\documentclass[12pt,prd,showpacs,tightenlines,nofootinbib]{revtex4}
\usepackage{bm}
\usepackage{graphics}
\usepackage{rotating}
\usepackage{epsfig}
\begin{document}
\title{\begin{flushright}{\rm\normalsize HU-EP-03/35}\end{flushright}
Weak decays of the $\bm{B_c}$ meson to charmonium and $\bm{D}$ mesons in the
relativistic quark model}  
\author{D. Ebert}
\affiliation{Institut f\"ur Physik, Humboldt--Universit\"at zu Berlin,
Newtonstr. 15, D-12489  Berlin, Germany}
\author{R. N. Faustov}
\author{V. O. Galkin}
\affiliation{Institut f\"ur Physik, Humboldt--Universit\"at zu Berlin,
Newtonstr. 15, D-12489 Berlin, Germany}
\affiliation{Russian Academy of Sciences, Scientific Council for
Cybernetics, Vavilov Street 40, Moscow 117333, Russia}
\begin{abstract}
Semileptonic and nonleptonic decays of the $B_c$ meson to charmonium
and $D$ mesons are studied in the framework of
the relativistic quark model. The decay form factors are explicitly
expressed through the overlap integrals of the meson wave functions
in the whole accessible kinematical range. The relativistic meson wave
functions are used for the calculation of the decay rates. The
obtained  results are compared with the predictions of other approaches.     
\end{abstract}

\maketitle

\section{Introduction}
\label{sec:intro}

The investigation of weak decays of mesons composed of a heavy quark and
antiquark gives a very important insight in the heavy quark dynamics.
The properties of the $B_c$ meson are of special interest, since it
is the only heavy meson consisting of two heavy quarks with different
flavor. This difference of quark flavors forbids annihilation into
gluons. As a result, the excited $B_c$ meson states lying below the
$BD$ production threshold undergo pionic or radiative transitions to
the pseudoscalar ground state which is considerably more
stable than corresponding charmonium or bottomonium states and  decays
only weakly.  The
Collider Detector at Fermilab (CDF) Collaboration \cite{cdfcol}
reported the discovery of the $B_c$ ground state in $p\bar p$
collisions. More experimental data are expected to come in near future
from the Tevatron and Large Hadron Collider (LHC).

The characteristic feature of the $B_c$ meson  is that both
quarks forming it are heavy and thus their weak decays give comparable
contributions to the total decay rate. Therefore it is necessary to
consider both the $b$ quark decays $b\to c,u$ with the $\bar c$ quark
being a spectator and $\bar c$ quark decays $\bar c\to \bar s,\bar d$
with $b$ quark being a
spectator. The former transitions lead to semileptonic decays to
charmonium and $D$ mesons while the latter lead to decays to $B_s$ and
$B$ mesons. 
The estimates of the $B_c$ decay rates indicate that the $c$ quark decays
give the dominant contribution ($\sim 70\%$) while the $b$ quark
decays and weak annihilation contribute about 20\% and 10\%,
respectively (for a recent review see e.g. \cite{gklry} and references
therein). However, from the experimental point of view the $B_c$
decays to charmonium are easier to identify. Indeed, CDF observed
$B_c$ mesons \cite{cdfcol} analysing their semileptonic
decays  $B_c\to J/\psi l\nu$. 

The important difference between the $B_c$ semileptonic decays induced by
$b\to c,u$ and $c\to s,d$ transitions lies in the substantial
difference of their kinematical ranges. In the case of $B_c$ decays to
charmonium and $D^{(*)}$ mesons the kinematical range (the square of 
momentum transfer to the lepton pair varies from 0 to $q^2_{\rm max}\approx
10$~GeV$^2$ for decays to $J/\psi$ and $q^2_{\rm max}\approx
18$~GeV$^2$ for decays to $D$
mesons)  is considerably broader than for decays to $B_s^{(*)}$  and
$B^{(*)}$ mesons ($q^2_{\rm max}\approx 0.8$~GeV$^2$ for decays to $B_s$ and
$q^2_{\rm max}\approx 1$~GeV$^2$ for decays to $B$ mesons). As a
result in the 
$B_c$ meson rest frame the maximum recoil momentum of the final
charmonium and $D$ mesons is of the same order of magnitude as their
masses, while the maximum recoil momentum of the $B_s^{(*)}$ and
$B^{(*)}$ mesons is considerably smaller than the meson masses.   
This significant difference in kinematics makes it reasonable to
consider $B_c$ decays induced by $b$ and $c$ quark decays separately.   

In this paper we consider weak $B_c$ decays to charmonium and
$D$ mesons in the framework of the relativistic quark model based on
the quasipotential approach in quantum field theory. This model has
been successfully applied for the calculations of mass spectra,
radiative and weak decays of heavy quarkonia and heavy-light mesons
\cite{mass1,mass,hlm,gf,fg,mod}. In our recent
paper \cite{efgbc} we applied this model for the investigation of
properties of the $B_c$ meson and heavy quarkonia. The relativistic
wave functions obtained there are used for the calculation of the
transition matrix elements. The consistent theoretical description of
$B_c$ decays 
to charmonium and $D$ mesons requires a reliable determination of the
$q^2$ dependence of the decay amplitudes in the whole
kinematical range. In most previous calculations the
corresponding decay form factors were determined only at one
kinematical point either $q^2=0$ or $q^2=q^2_{\rm max}$ and then
extrapolated to the allowed kinematical range using some
phenomenological ansatz (mainly (di)pole or Gaussian). Our aim is to
explicitly  determine the $q^2$ dependence of form factors in the
whole kinematical range in order to avoid extrapolations thus reducing
uncertainties. The large values of recoil momentum require the
consistent relativistic treatment of these decays. In particular,
the relativistic transformation of the meson wave functions from the
moving to the rest reference frame should be taken into account. On
the other hand, the presence of only heavy quarks  in $B_c$
and charmonium allows one to use expansions in the inverse powers of
heavy quark masses $1/m_{b,c}$.  

The paper is organized as follows. In Sec.~\ref{rqm} we describe the
underlying 
relativistic quark model. The method for calculating matrix elements
of the weak current for $b\to c,u$ transitions in $B_c$ meson decays
is presented in Sec.~\ref{mml}. Special attention is devoted to the
dependence of the decay amplitudes on the momentum transfer. The $B_c$
decay form factors are calculated in the whole kinematical range in
Sec.~\ref{dff}. The $q^2$ dependence of the form factors is explicitly
determined. These form factors are used for the calculation of the
$B_c$ semileptonic decay rates in Sec.~\ref{ssd}. Section~\ref{nl}
contains our predictions for the energetic nonleptonic $B_c$ decays in the
factorization approximation, and a comparison of our results with other
theoretical calculations is presented. Our conclusions are given in
Sec.~\ref{sec:conc}. Finally, the Appendix contains complete 
expressions for the decay form factors.

\section{Relativistic quark model}  
\label{rqm}

In the quasipotential approach a meson is described by the wave
function of the bound quark-antiquark state, which satisfies the
quasipotential equation \cite{3} of the Schr\"odinger type \cite{4}
\begin{equation}
\label{quas}
{\left(\frac{b^2(M)}{2\mu_{R}}-\frac{{\bf
p}^2}{2\mu_{R}}\right)\Psi_{M}({\bf p})} =\int\frac{d^3 q}{(2\pi)^3}
 V({\bf p,q};M)\Psi_{M}({\bf q}),
\end{equation}
where the relativistic reduced mass is
\begin{equation}
\mu_{R}=\frac{E_1E_2}{E_1+E_2}=\frac{M^4-(m^2_1-m^2_2)^2}{4M^3},
\end{equation}
and $E_1$, $E_2$ are the center of mass energies on mass shell given by
\begin{equation}
\label{ee}
E_1=\frac{M^2-m_2^2+m_1^2}{2M}, \quad E_2=\frac{M^2-m_1^2+m_2^2}{2M}.
\end{equation}
Here $M=E_1+E_2$ is the meson mass, $m_{1,2}$ are the quark masses,
and ${\bf p}$ is their relative momentum.  
In the center of mass system the relative momentum squared on mass shell 
reads
\begin{equation}
{b^2(M) }
=\frac{[M^2-(m_1+m_2)^2][M^2-(m_1-m_2)^2]}{4M^2}.
\end{equation}

The kernel 
$V({\bf p,q};M)$ in Eq.~(\ref{quas}) is the quasipotential operator of
the quark-antiquark interaction. It is constructed with the help of the
off-mass-shell scattering amplitude, projected onto the positive
energy states. 
Constructing the quasipotential of the quark-antiquark interaction, 
we have assumed that the effective
interaction is the sum of the usual one-gluon exchange term with the mixture
of long-range vector and scalar linear confining potentials, where
the vector confining potential
contains the Pauli interaction. The quasipotential is then defined by
\cite{mass1}
  \begin{equation}
\label{qpot}
V({\bf p,q};M)=\bar{u}_1(p)\bar{u}_2(-p){\mathcal V}({\bf p}, {\bf
q};M)u_1(q)u_2(-q),
\end{equation}
with
$${\mathcal V}({\bf p},{\bf q};M)=\frac{4}{3}\alpha_sD_{ \mu\nu}({\bf
k})\gamma_1^{\mu}\gamma_2^{\nu}
+V^V_{\rm conf}({\bf k})\Gamma_1^{\mu}
\Gamma_{2;\mu}+V^S_{\rm conf}({\bf k}),$$
where $\alpha_s$ is the QCD coupling constant, $D_{\mu\nu}$ is the
gluon propagator in the Coulomb gauge
\begin{equation}
D^{00}({\bf k})=-\frac{4\pi}{{\bf k}^2}, \quad D^{ij}({\bf k})=
-\frac{4\pi}{k^2}\left(\delta^{ij}-\frac{k^ik^j}{{\bf k}^2}\right),
\quad D^{0i}=D^{i0}=0,
\end{equation}
and ${\bf k=p-q}$; $\gamma_{\mu}$ and $u(p)$ are 
the Dirac matrices and spinors
\begin{equation}
\label{spinor}
u^\lambda({p})=\sqrt{\frac{\epsilon(p)+m}{2\epsilon(p)}}
\left(
\begin{array}{c}1\cr {\displaystyle\frac{\bm{\sigma}
      {\bf  p}}{\epsilon(p)+m}}
\end{array}\right)\chi^\lambda.
\end{equation}
Here  $\bm{\sigma}$   and $\chi^\lambda$
are the Pauli matrices and spinors; $\epsilon(p)=\sqrt{p^2+m^2}$.
The effective long-range vector vertex is
given by
\begin{equation}
\label{kappa}
\Gamma_{\mu}({\bf k})=\gamma_{\mu}+
\frac{i\kappa}{2m}\sigma_{\mu\nu}k^{\nu},
\end{equation}
where $\kappa$ is the Pauli interaction constant characterizing the
long-range anomalous chromomagnetic moment of quarks. Vector and
scalar confining potentials in the nonrelativistic limit reduce to
\begin{eqnarray}
\label{vlin}
V_V(r)&=&(1-\varepsilon)Ar+B,\nonumber\\ 
V_S(r)& =&\varepsilon Ar,
\end{eqnarray}
reproducing 
\begin{equation}
\label{nr}
V_{\rm conf}(r)=V_S(r)+V_V(r)=Ar+B,
\end{equation}
where $\varepsilon$ is the mixing coefficient. 

The expression for the quasipotential of the heavy quarkonia,
expanded in $v^2/c^2$ without and with retardation corrections to the
confining potential, can be found in Refs.~\cite{mass1} and \cite{efgbc,mass},
respectively. The 
structure of the spin-dependent interaction is in agreement with
the parameterization of Eichten and Feinberg \cite{ef}. The
quasipotential for the heavy quark interaction with a light antiquark
without employing the expansion in inverse powers of the light quark
mass is given in Ref.~\cite{hlm}.  
All the parameters of
our model like quark masses, parameters of the linear confining potential
$A$ and $B$, mixing coefficient $\varepsilon$ and anomalous
chromomagnetic quark moment $\kappa$ are fixed from the analysis of
heavy quarkonium masses \cite{mass1} and radiative
decays \cite{gf}. The quark masses
$m_b=4.88$ GeV, $m_c=1.55$ GeV, $m_{u,d}=0.33$ GeV and
the parameters of the linear potential $A=0.18$ GeV$^2$ and $B=-0.16$ GeV
have usual values of quark models.  The value of the mixing
coefficient of vector and scalar confining potentials $\varepsilon=-1$
has been determined from the consideration of the heavy quark expansion
for the semileptonic $B\to D$ decays
\cite{fg} and charmonium radiative decays \cite{gf}.
Finally, the universal Pauli interaction constant $\kappa=-1$ has been
fixed from the analysis of the fine splitting of heavy quarkonia ${
}^3P_J$- states \cite{mass1}. Note that the 
long-range  magnetic contribution to the potential in our model
is proportional to $(1+\kappa)$ and thus vanishes for the 
chosen value of $\kappa=-1$. It has been known for a long time 
that the correct reproduction of the
spin-dependent part of the quark-antiquark interaction requires 
either assuming  the scalar confinement or equivalently  introducing the
Pauli interaction with $\kappa=-1$ \cite{schn,mass1,mass} in the vector
confinement.

\section{Matrix elements of the electroweak current for
  $\bm{\lowercase{b\to c,u}}$ transitions} \label{mml}

In order to calculate the exclusive semileptonic decay rate of the
$B_c$ meson, it is necessary to determine the corresponding matrix
element of the  weak current between meson states.
In the quasipotential approach,  the matrix element of the weak current
$J^W_\mu=\bar q\gamma_\mu(1-\gamma_5)b$, associated with $b\to q$ ($q=c$
or $u$) transition, between a $B_c$ meson with mass $M_{B_c}$ and
momentum $p_{B_c}$ and a final meson $F$ ($F=\psi,\eta_c$ or
$D^{(*)}$) with mass $M_F$ and momentum $p_F$ takes the form \cite{f}
\begin{equation}\label{mxet} 
\langle F(p_F) \vert J^W_\mu \vert B_c(p_{B_c})\rangle
=\int \frac{d^3p\, d^3q}{(2\pi )^6} \bar \Psi_{F\,{\bf p}_F}({\bf
p})\Gamma _\mu ({\bf p},{\bf q})\Psi_{B_c\,{\bf p}_{B_c}}({\bf q}),
\end{equation}
where $\Gamma _\mu ({\bf p},{\bf
q})$ is the two-particle vertex function and  
$\Psi_{M\,{\bf p}_M}$ are the
meson ($M=B_c,F)$ wave functions projected onto the positive energy states of
quarks and boosted to the moving reference frame with momentum ${\bf p}_M$.
\begin{figure}
  \centering
  \includegraphics{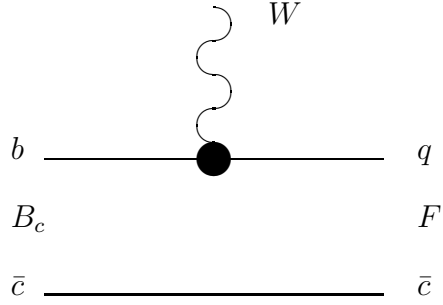}
\caption{Lowest order vertex function $\Gamma^{(1)}$
contributing to the current matrix element (\ref{mxet}). \label{d1}}
\end{figure}

\begin{figure}
  \centering
  \includegraphics{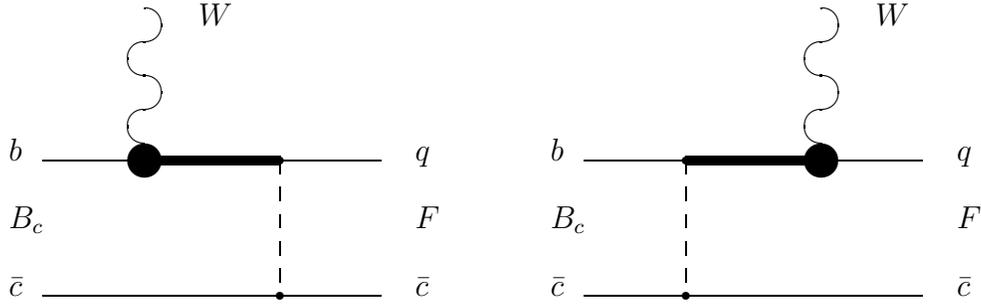}
\caption{ Vertex function $\Gamma^{(2)}$
taking the quark interaction into account. Dashed lines correspond  
to the effective potential ${\cal V}$ in 
(\ref{qpot}). Bold lines denote the negative-energy part of the quark
propagator. \label{d2}}
\end{figure}

 The contributions to $\Gamma$ come from Figs.~\ref{d1} and \ref{d2}. 
The contribution $\Gamma^{(2)}$ is the consequence
of the projection onto the positive-energy states. Note that the form of the
relativistic corrections resulting from the vertex function
$\Gamma^{(2)}$ is explicitly dependent on the Lorentz structure of the
quark-antiquark interaction. In the leading order of the $v^2/c^2$
expansion for 
$B_c$ and $\psi$ and in the heavy quark limit $m_{c}\to \infty$ for $D$
only $\Gamma^{(1)}$ contributes, while $\Gamma^{(2)}$  
contributes already at the subleading order. 
The vertex functions look like
\begin{equation} \label{gamma1}
\Gamma_\mu^{(1)}({\bf
p},{\bf q})=\bar u_{q}(p_q)\gamma_\mu(1-\gamma^5)u_b(q_b)
(2\pi)^3\delta({\bf p}_c-{\bf
q}_c),\end{equation}
and
\begin{eqnarray}\label{gamma2} 
\Gamma_\mu^{(2)}({\bf
p},{\bf q})&=&\bar u_{q}(p_q)\bar u_c(p_c) \Bigl\{\gamma_{1\mu}(1-\gamma_1^5)
\frac{\Lambda_b^{(-)}(
k)}{\epsilon_b(k)+\epsilon_b(p_q)}\gamma_1^0
{\cal V}({\bf p}_c-{\bf
q}_c)\nonumber \\ 
& &+{\cal V}({\bf p}_c-{\bf
q}_c)\frac{\Lambda_{q}^{(-)}(k')}{ \epsilon_{q}(k')+
\epsilon_{q}(q_b)}\gamma_1^0 \gamma_{1\mu}(1-\gamma_1^5)\Bigr\}u_b(q_b)
u_c(q_c),\end{eqnarray}
where the superscripts ``(1)" and ``(2)" correspond to Figs.~\ref{d1} and
\ref{d2},  ${\bf k}={\bf p}_q-{\bf\Delta};\
{\bf k}'={\bf q}_b+{\bf\Delta};\ {\bf\Delta}={\bf
p}_F-{\bf p}_{B_c}$;
$$\Lambda^{(-)}(p)=\frac{\epsilon(p)-\bigl( m\gamma
^0+\gamma^0({\bm{ \gamma}{\bf p}})\bigr)}{ 2\epsilon (p)}.$$
Here \cite{f} 
\begin{eqnarray*} 
p_{q,c}&=&\epsilon_{q,c}(p)\frac{p_F}{M_F}
\pm\sum_{i=1}^3 n^{(i)}(p_F)p^i,\\
q_{b,c}&=&\epsilon_{b,c}(q)\frac{p_{B_c}}{M_{B_c}} \pm \sum_{i=1}^3 n^{(i)}
(p_{B_c})q^i,\end{eqnarray*}
and $n^{(i)}$ are three four-vectors given by
$$ n^{(i)\mu}(p)=\left\{ \frac{p^i}{M},\ \delta_{ij}+
\frac{p^ip^j}{M(E+M)}\right\}, \quad E=\sqrt{{\bf p}^2+M^2}.$$

It is important to note that the wave functions entering the weak current
matrix element (\ref{mxet}) are not in the rest frame in general. For example, 
in the $B_c$ meson rest frame (${\bf p}_{B_c}=0$), the final  meson
is moving with the recoil momentum ${\bf \Delta}$. The wave function
of the moving  meson $\Psi_{F\,{\bf\Delta}}$ is connected 
with the  wave function in the rest frame 
$\Psi_{F\,{\bf 0}}\equiv \Psi_F$ by the transformation \cite{f}
\begin{equation}
\label{wig}
\Psi_{F\,{\bf\Delta}}({\bf
p})=D_q^{1/2}(R_{L_{\bf\Delta}}^W)D_c^{1/2}(R_{L_{
\bf\Delta}}^W)\Psi_{F\,{\bf 0}}({\bf p}),
\end{equation}
where $R^W$ is the Wigner rotation, $L_{\bf\Delta}$ is the Lorentz boost
from the meson rest frame to a moving one, and   
the rotation matrix $D^{1/2}(R)$ in spinor representation is given by
\begin{equation}\label{d12}
{1 \ \ \,0\choose 0 \ \ \,1}D^{1/2}_{q,c}(R^W_{L_{\bf\Delta}})=
S^{-1}({\bf p}_{q,c})S({\bf\Delta})S({\bf p}),
\end{equation}
where
$$
S({\bf p})=\sqrt{\frac{\epsilon(p)+m}{2m}}\left(1+\frac{\bm{\alpha}{\bf p}}
{\epsilon(p)+m}\right)
$$
is the usual Lorentz transformation matrix of the four-spinor.

The general structure of the current matrix element (\ref{mxet}) is
rather complicated, because it is necessary to integrate both with
respect to $d^3p$ and $d^3q$. The $\delta$-function in the expression
(\ref{gamma1}) for the vertex function $\Gamma^{(1)}$ permits to perform
one of these integrations. As a result the contribution of
$\Gamma^{(1)}$ to the current matrix element has the usual structure of
an overlap integral of meson wave functions and
can be calculated exactly (without employing any expansion) in the
whole kinematical range, if the wave functions of the
initial and final meson are known. The situation with the contribution
$\Gamma^{(2)}$ is different. Here, instead of a $\delta$-function, we have
a complicated structure, containing the potential of the $q\bar
q$-interaction in meson. Thus in the general case we cannot get rid of one
of the integrations in the contribution of $\Gamma^{(2)}$ to the
matrix element (\ref{mxet}). Therefore, it is necessary to use some 
additional considerations in order to simplify calculations. The main
idea is to expand the vertex 
function $\Gamma^{(2)}$, given by (\ref{gamma2}), in such  a way that it
will be possible to use the quasipotential equation (\ref{quas}) in order
to perform one of the integrations in the current matrix element
(\ref{mxet}).  

\begin{figure}
  \centering
  \includegraphics[scale=1]{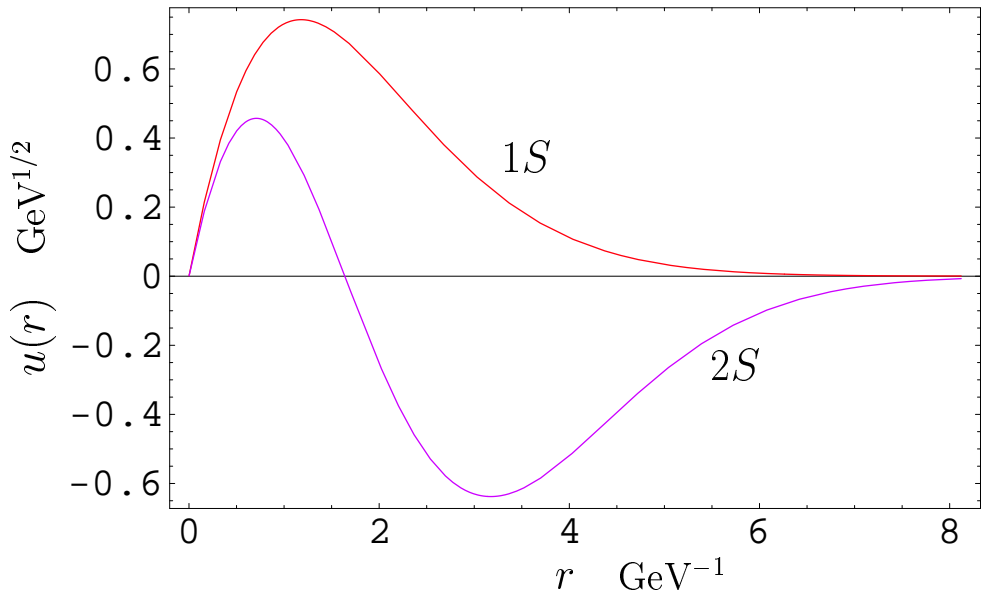}
\caption{The reduced radial wave functions for the $S$-states of the
  $B_c$ meson. \label{bcwf}} 
\end{figure}

{ \bf  $\bm{B_c\to \psi,\eta_c}$ transitions.} The
natural expansion parameters for $B_c$ decays to charmonium are the
active heavy $b$ and $c$ quark masses as well as the spectator $c$
quark mass. We carry such an expansion up to the second order in the ratios
of the relative quark momentum ${\bf p}$ and binding energy to the
heavy quark masses $m_{b,c}$. It is important to take into account the
fact that in the case of weak $B_c$ decays caused by $b\to c,u$ quark
transition the kinematically allowed range is large
($|{\bf\Delta}_{\rm max}|=(M_{B_c}^2-M_F^2)/(2M_{B_c})$ $\sim 2.4$~GeV for
decays to charmonium and $\sim 2.8$~GeV for decays to $D$ mesons). This
means that the recoil momentum ${\bf \Delta}$ of a final meson is large in
comparison to the relative momentum ${\bf p}$ of quarks inside a meson 
($\sim 0.5$~GeV), being of the same order as the heavy
quark mass  almost in the whole kinematical range. Thus we do not use
expansions in powers of $|{\bf\Delta}|/m_{b,c}$ or
$|{\bf\Delta}|/M_F$, but approximate in the expression (\ref{gamma2}) for
$\Gamma^{(2)}$ the heavy
quark energies $\epsilon_{b,c}(p+\Delta)\equiv\sqrt{m_{b,c}^2+({\bf
p}+{\bf\Delta})^2}$ by $\epsilon_{b,c}(\Delta)\equiv
\sqrt{m_{b,c}^2+{\bf\Delta}^2}$, which become independent of the quark
relative momentum 
${\bf p}$. Making these replacements and expansions we see that it is
possible to 
integrate the current matrix element (\ref{mxet}) either with
respect to $d^3p$ or $d^3q$ using the quasipotential equation
(\ref{quas}). Performing integrations and
taking the sum of the contributions of $\Gamma^{(1)}$ and
$\Gamma^{(2)}$ we get the expression for the current matrix element,
which contains ordinary overlap integrals of meson wave functions
and is valid in the whole kinematical range.
Thus this matrix element can be easily calculated using numerical wave
functions found in our meson mass spectrum analysis \cite{efgbc,mass}. 
The reduced radial wave functions $u(r)\equiv rR(r)$ of the $B_c$
meson are shown in Fig.~\ref{bcwf}.  

{ \bf  $\bm{B_c\to D^{(*)}}$ transitions.} In this case the heavy
$b$ quark undergoes the weak transition to the light $u$ quark. The
constituent $u$ quark mass is of the same order of magnitude as the relative
momentum and binding energy, thus we cannot apply the expansion in
inverse powers of its mass. Nevertheless, taking into account the fact
that the recoil momentum of the final meson in this decay is large
almost in the whole kinematical range (as it was discussed above), we
can neglect the relative momentum ${\bf p}$ of quarks inside a meson
with respect to the large recoil momentum ${\bf \Delta}$. Thus in the
region of large recoil ($|{\bf \Delta}|\gg |{\bf p}| $) we can
use the same expressions of the $\Gamma^{(2)}$ contribution to the
current matrix element both for the $B_c\to D^{(*)}$ and
$B_c\to\psi,\eta_c$  transitions. Moreover, the smallness of
the $\Gamma^{(2)}$ contribution, 
which is proportional to the small binding energy, and its weak
dependence on momentum transfer allows one to extrapolate these
formulae to the whole kinematical range. As numerical estimates show
(see below), such extrapolation introduces only small uncertainties.

\section{$\bm{B_{\lowercase{c}}}$ decay form factors}\label{dff}

 The matrix elements of the weak current $J^W$ for $B_c$ decays
 to pseudoscalar  mesons
($P=\eta_c, D$) can be parametrized by two invariant form factors:
\begin{equation}
  \label{eq:pff1}
  \langle P(p_F)|\bar q \gamma^\mu b|B_c(p_{B_c})\rangle
  =f_+(q^2)\left[p_{B_c}^\mu+ p_F^\mu-
\frac{M_{B_c}^2-M_P^2}{q^2}\ q^\mu\right]+
  f_0(q^2)\frac{M_{B_c}^2-M_P^2}{q^2}\ q^\mu,
\end{equation}
where $q=p_{B_c}-p_F$; $M_{B_c}$ is the $B_c$ meson mass 
and $M_P$ is the pseudoscalar meson mass. 

The corresponding matrix elements for $B_c$  decays to vector mesons
($V=J/\psi, D^*$) are parametrized by four form factors
\begin{eqnarray}
  \label{eq:vff1}
  \langle V(p_F)|\bar q \gamma^\mu b|B(p_{B_c})\rangle&=
  &\frac{2iV(q^2)}{M_{B_c}+M_V} \epsilon^{\mu\nu\rho\sigma}\epsilon^*_\nu
  p_{B_c\rho} p_{F\sigma},\\ \cr
\label{eq:vff2}
\langle V(p_F)|\bar q \gamma^\mu\gamma_5 b|B(p_{B_c})\rangle&=&2M_V
A_0(q^2)\frac{\epsilon^*\cdot q}{q^2}\ q^\mu
 +(M_{B_c}+M_V)A_1(q^2)\left(\epsilon^{*\mu}-\frac{\epsilon^*\cdot
    q}{q^2}\ q^\mu\right)\cr\cr
&&-A_2(q^2)\frac{\epsilon^*\cdot q}{M_{B_c}+M_V}\left[p_{B_c}^\mu+
  p_F^\mu-\frac{M_{B_c}^2-M_V^2}{q^2}\ q^\mu\right], 
\end{eqnarray}
where 
$M_V$ and $\epsilon_\mu$ are the mass and polarization vector of
the final vector meson. The following relations hold for the form
factors at the maximum recoil point of the final meson ($q^2=0$)
\[f_+(0)=f_0(0),\]
\[A_0(0)=\frac{M_{B_c}+M_V}{2M_V}A_1(0)
-\frac{M_{B_c}-M_V}{2M_V}A_2(0).\]
In the limit of vanishing lepton mass, the form factors $f_0$ and $A_0$ do
not contribute to the semileptonic decay rates. However, they
contribute to nonleptonic decay rates in the factorization
approximation.  

It is convenient to consider $B_c$ semileptonic and nonleptonic decays
in the $B_c$ meson rest frame. Then it is important to take into account 
the boost of the final meson wave function from the rest reference
frame to the moving one with the recoil momentum ${\bf \Delta}$, given
by Eq.~(\ref{wig}). Now we can apply the method for
calculating decay matrix elements described in the previous section.   
As it is argued above, the leading contributions arising from the vertex
function $\Gamma^{(1)}$ can be exactly expressed through the overlap
integrals of the meson wave functions in the whole kinematical range. 
For the subleading contribution $\Gamma^{(2)}$, the expansion in powers
of the ratio of the relative quark momentum ${\bf p}$ to heavy quark
masses $m_{b,c}$ should be performed taking into account that the
recoil momentum of the final meson ${\bf \Delta}$ can be large. Such
expansion is well justified for $B_c$ decays to charmonium in the
whole kinematical range. For $B_c$ decays to $D$ mesons, where 
one of the final quarks is light, a similar expansion is well
justified only in the kinematical region of large recoil
momentum. However, the numerical smallness of this subleading
contribution due to its proportionality to the small meson
binding energy permits its extrapolation to the whole kinematical range.
As a result, we get the following expressions for the $B_c$ decay
form factors:

(a) $B_c\to P$ transitions ($P=\eta_c,D$) 
\begin{equation}
  \label{eq:f+}
  f_+(q^2)=f_+^{(1)}(q^2)+\varepsilon f_+^{S(2)}(q^2)
+(1-\varepsilon) f_+^{V(2)}(q^2),
\end{equation}
\begin{equation}
  \label{eq:f0}
  f_0(q^2)=f_0^{(1)}(q^2)+\varepsilon f_0^{S(2)}(q^2)
+(1-\varepsilon) f_0^{V(2)}(q^2),
\end{equation}

(b) $B_c\to V$ transition ($V=\psi,D^*$)
\begin{equation}
  \label{eq:V}
  V(q^2)=V^{(1)}(q^2)+\varepsilon V^{S(2)}(q^2)
+(1-\varepsilon) V^{V(2)}(q^2),
\end{equation}
\begin{equation}
  \label{eq:A1}
  A_1(q^2)=A_1^{(1)}(q^2)+\varepsilon A_1^{S(2)}(q^2)
+(1-\varepsilon) A_1^{V(2)}(q^2),
\end{equation}
\begin{equation}
  \label{eq:A2}
  A_2(q^2)=A_2^{(1)}(q^2)+\varepsilon A_2^{S(2)}(q^2)
+(1-\varepsilon) A_2^{V(2)}(q^2),
\end{equation}
\begin{equation}
  \label{eq:A0}
  A_0(q^2)=A_0^{(1)}(q^2)+\varepsilon A_0^{S(2)}(q^2)
+(1-\varepsilon) A_0^{V(2)}(q^2),
\end{equation}
where $f_{+,0}^{(1)}$, $f_{+,0}^{S,V(2)}$, $A_{0,1,2}^{(1)}$,
$A_{0,1,2}^{S,V(2)}$, $V^{(1)}$ and $V^{S,V(2)}$ are given in Appendix.
The superscripts ``(1)" and ``(2)" correspond to Figs.~\ref{d1} and
\ref{d2}, $S$ and
$V$ to the scalar and vector potentials of $q\bar q$-interaction.
The mixing parameter of scalar and vector confining potentials
$\varepsilon$ is fixed to be $-1$  in our model.

It is easy to check that in the heavy quark limit the decay matrix
elements (\ref{eq:pff1})--(\ref{eq:vff2}) with form factors
(\ref{eq:f+})--(\ref{eq:A0}) satisfy the heavy quark spin symmetry
relations \cite{jlms} obtained near the zero recoil point (${\bf
  \Delta}\to 0$).   

\begin{table}
\caption{Form factors of weak $B_c$ decays ($b\to c,u$ transitions). }
\label{ff}
\begin{ruledtabular}
\begin{tabular}{ccccccc}
Transition   & $f_+(q^2)$ & $f_0(q^2)$ & $V(q^2)$ & $A_1(q^2)$ & $A_2(q^2)$ &
   $A_0(q^2)$\\
\hline
$B_c\to\eta_c,J/\psi$ & \\
$q^2=q^2_{\rm max}$ & 1.07 & 0.92 & 1.34 & 0.88 & 1.33 & 1.06\\  
$q^2=0$ & 0.47 & 0.47 & 0.49 & 0.50 & 0.73 & 0.40\\
$B_c\to\eta_c',\psi'$ & \\
$q^2=q^2_{\rm max}$ & 0.08 & 0.05 & $-0.16$ & 0.03 & 0.10 & 0.08\\
$q^2=0$ & 0.27 & 0.27 & 0.24 & 0.18 & 0.14 & 0.23\\
$B_c\to D,D^*$ &\\
$q^2=q^2_{\rm max}$ & 1.20 & 0.64 & 2.60 & 0.62 & 1.78 & 0.97   \\
$q^2=0$ & 0.14 & 0.14 & 0.18 & 0.17 & 0.19 & 0.14 \\
\end{tabular}
\end{ruledtabular}
\end{table}

\begin{figure}
  \centering
 \includegraphics{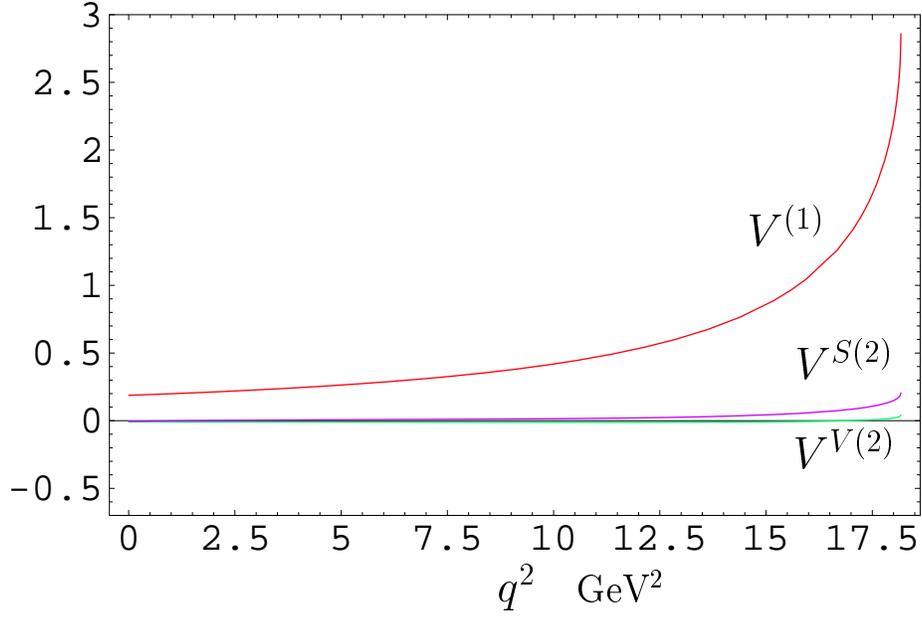}
\caption{Leading $V^{(1)}$ and subleading $V^{S(2)}$, $V^{V(2)}$
  contributions to the form factor $V$ for the $B_c\to D^*$ transition.}
  \label{fvbcd}  
\end{figure}

For numerical calculations we use the quasipotential wave functions of the
$B_c$ meson, charmonium and $D$ mesons obtained in the mass spectra
calculations \cite{mass,hlm}.  Our model predicts the $B_c$ meson mass  
$M_{B_c}=6.270$~GeV \cite{efgbc}, while for $J/\psi$, $\eta_c$, $\psi'$,
$\eta_c'$, $D$ and $D^*$ meson masses we use experimental data \cite{pdg}.
The calculated values of form factors at zero 
($q^2=q^2_{\rm max}$) and maximum ($q^2=0$) recoil of the final meson
are listed in Table~\ref{ff}. In Fig.~\ref{fvbcd} we plot leading
$V^{(1)}$ and subleading $V^{S(2)}$, $V^{V(2)}$ contributions to the
form factor $V$ for $B_c\to D^*$ transition, as an example. We see
that the leading contribution $V^{(1)}$ is dominant in the whole
kinematical range, as it was expected. The subleading contributions
$V^{S(2)}$, $V^{V(2)}$ are small and depend weakly on $q^2$. The
behavior of corresponding contributions to other form factors is
similar. This supports our conjecture that the formulae 
(\ref{eq:fpl})--(\ref{eq:a0v}) can be applied for the calculation of the form
factors of $B_c\to D^{(*)}$ transitions in the whole kinematical
range. 

\begin{figure}
  \centering
 \includegraphics{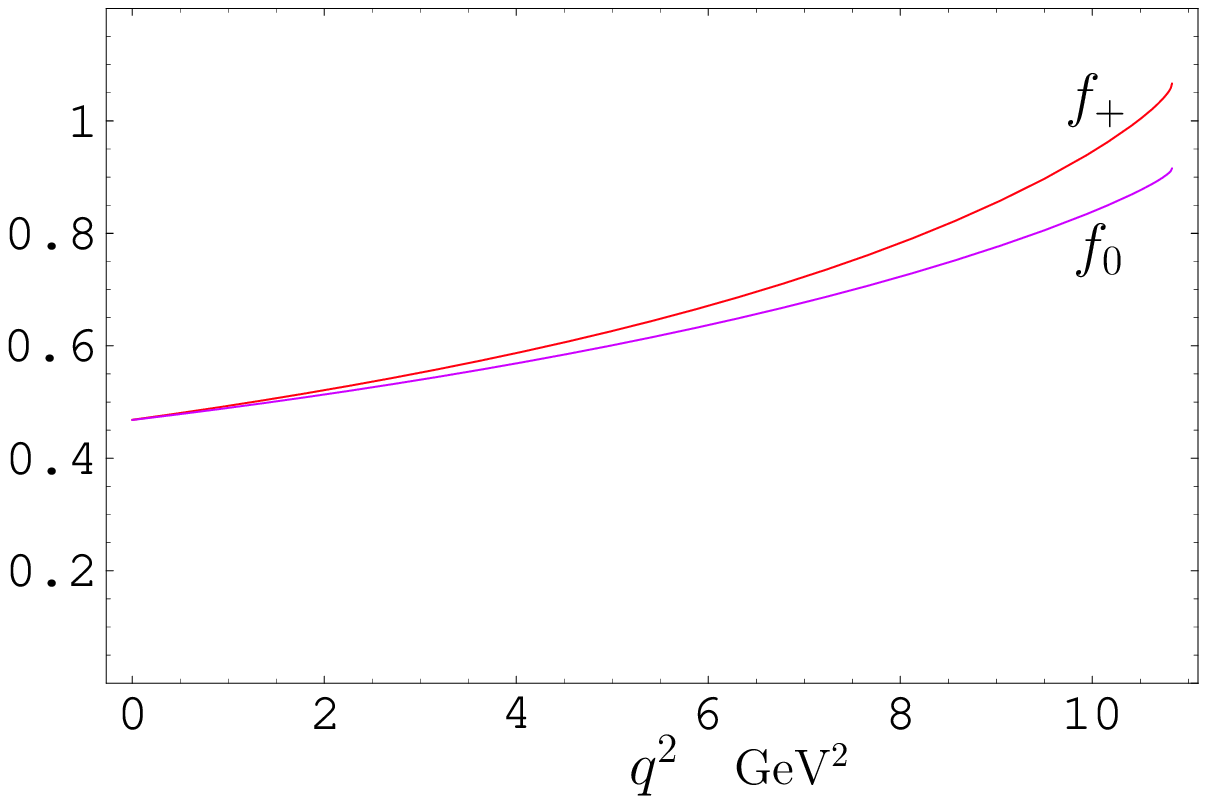}
\caption{Form factors of $B_c\to\eta_c e\nu$ decay.
\label{bctoetacff}} 
\end{figure}

\begin{figure}
  \centering
\includegraphics{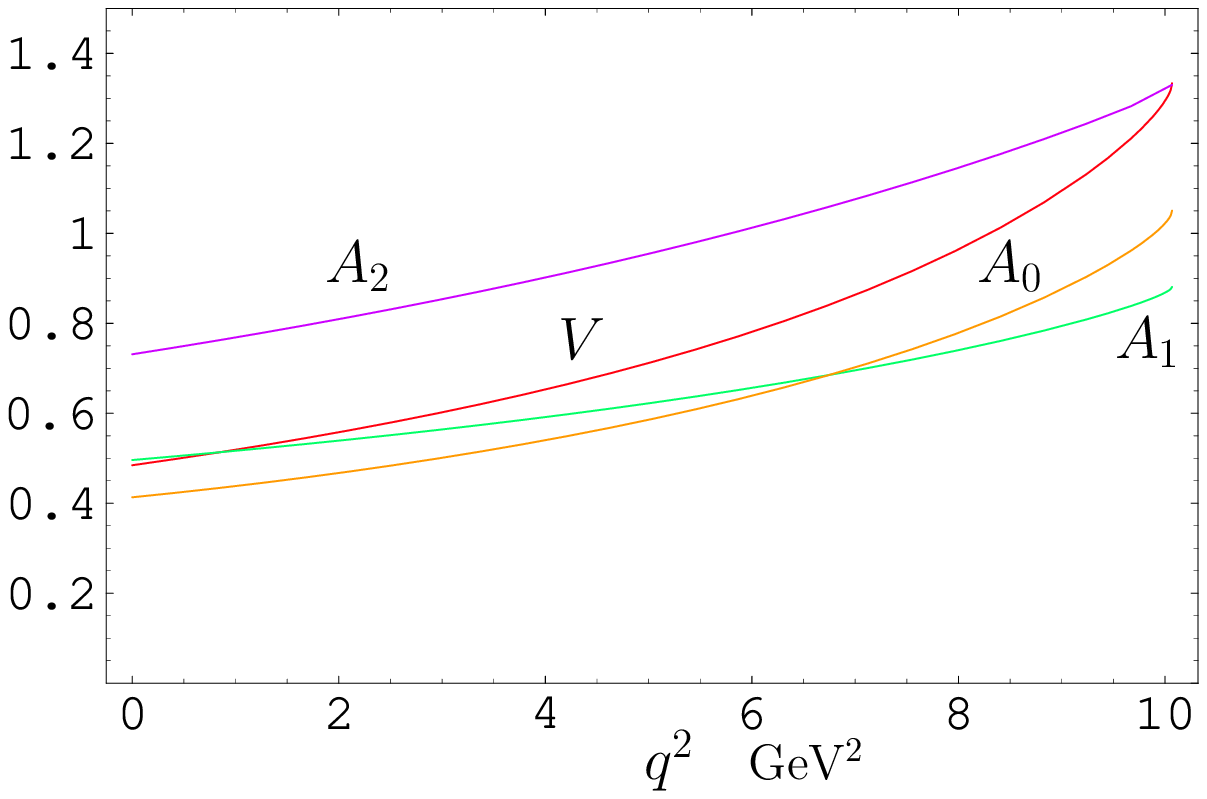}
\caption{Form factors of $B_c\to J/\psi e\nu$ decay.
\label{bctopsiff1}}
\end{figure}

\begin{figure}
  \centering
\includegraphics{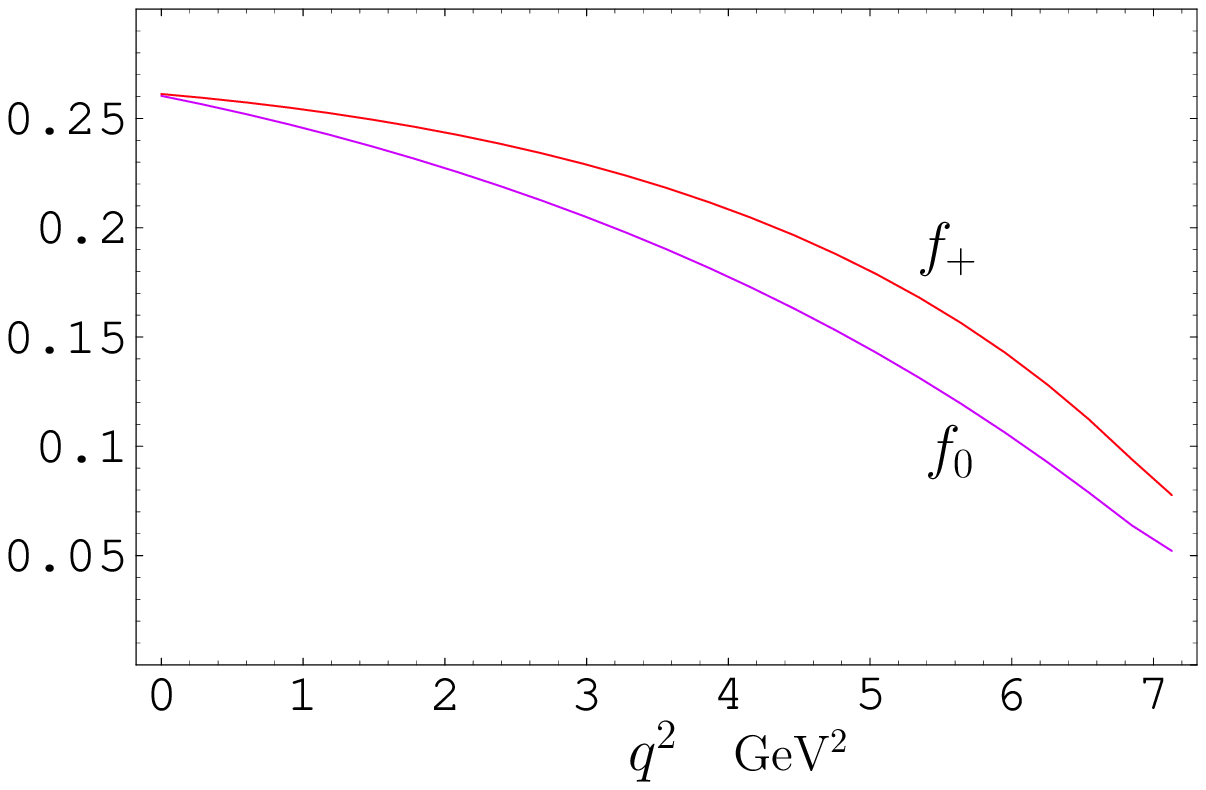}
\caption{Form factors of $B_c\to\eta_c' e\nu$ decay.
\label{bctoetaexcf}}
\end{figure}

\begin{figure}
  \centering
  \includegraphics{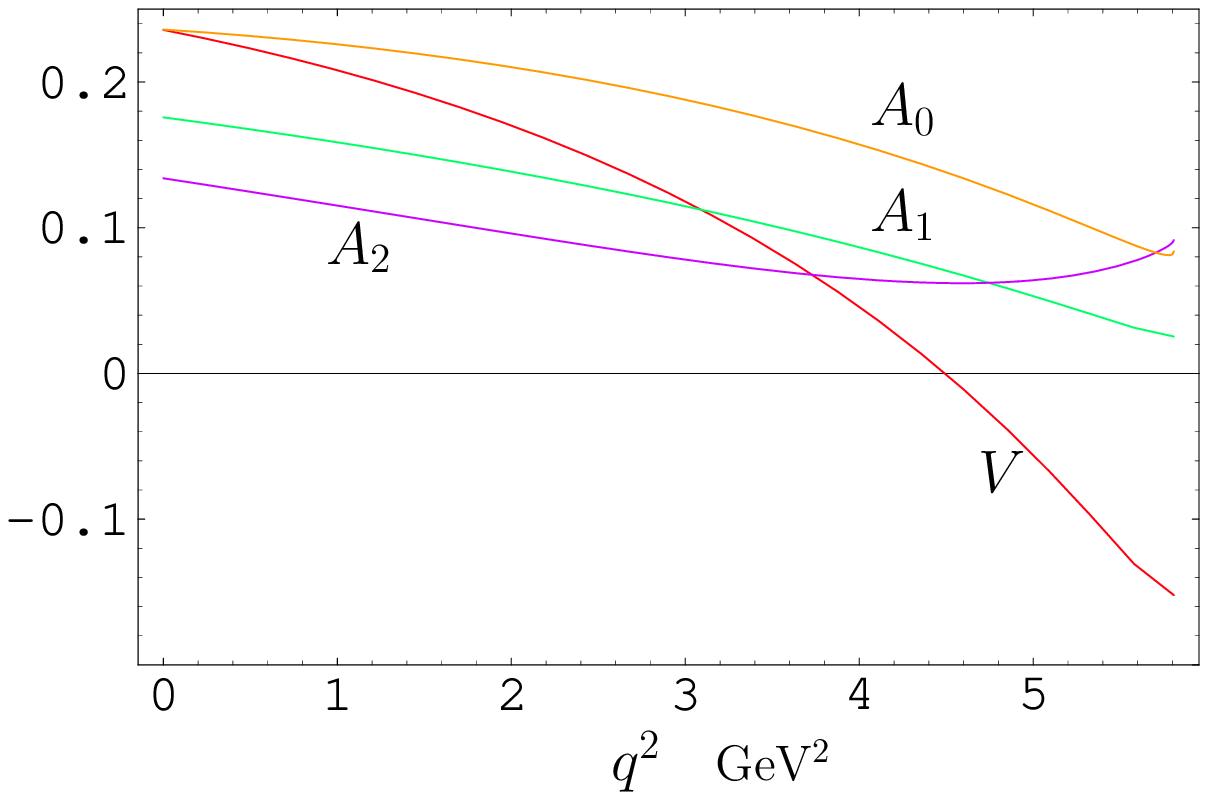}
\caption{Form factors of $B_c\to \psi' e\nu$ decay.
\label{bctopsiexcff}}
\end{figure}

\begin{figure}
  \centering
  \includegraphics{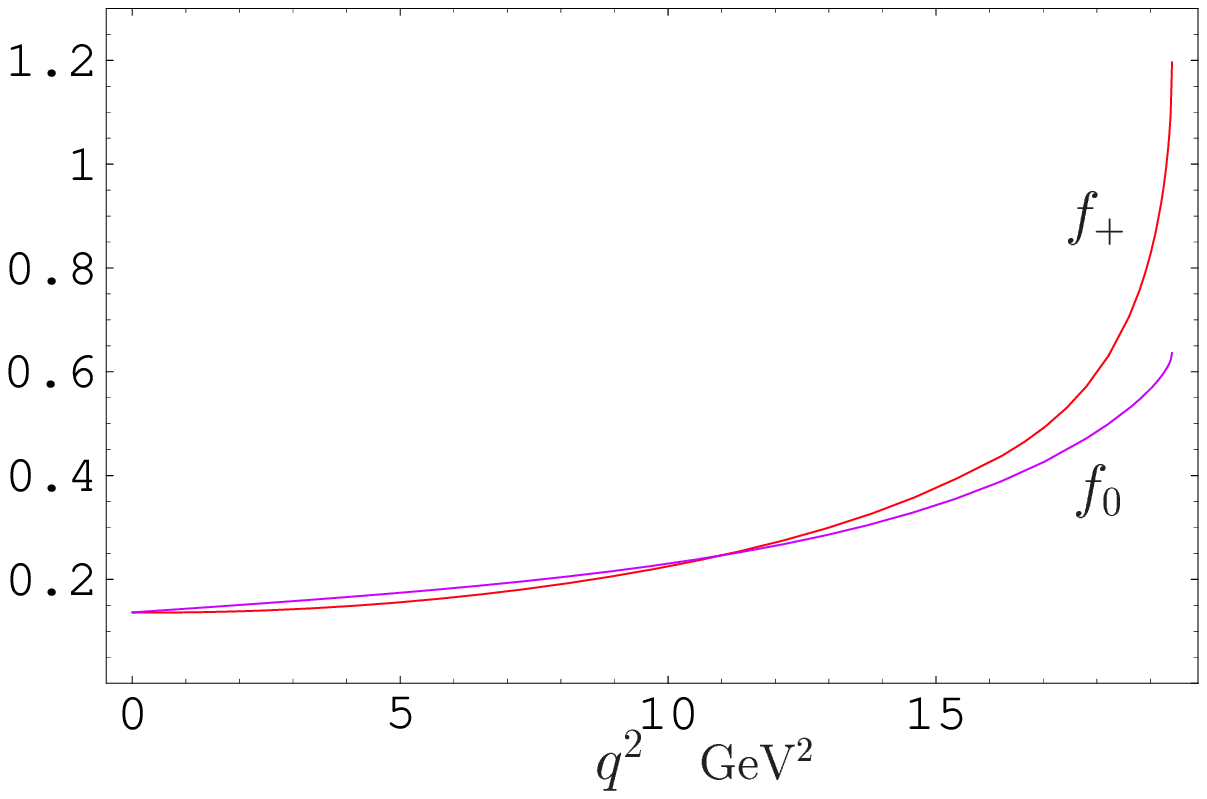}
\caption{Form factors of $B_c\to D e\nu$ decay.\label{bctodff}}
\end{figure}

\begin{figure}
  \centering
 \includegraphics{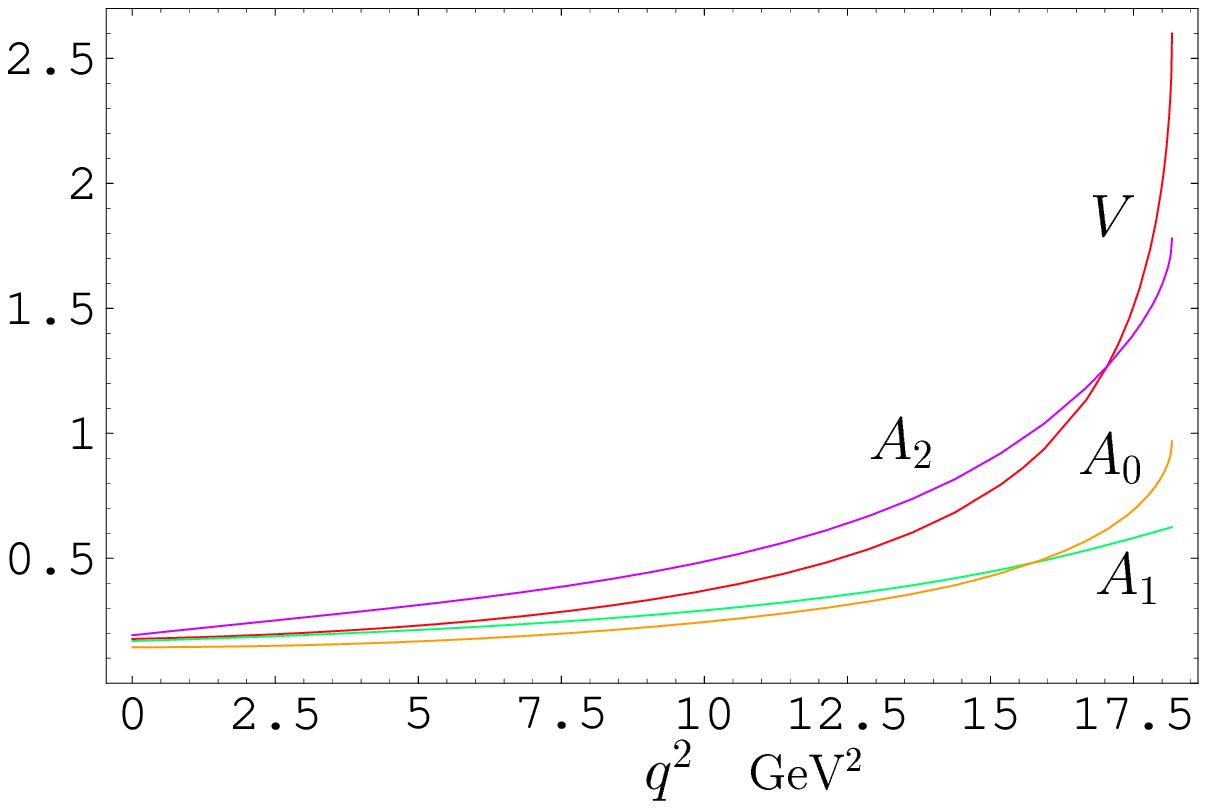}
\caption{Form factors of $B_c\to D^* e\nu$ decay.\label{bctodvff}}
\end{figure}

In Figs.~\ref{bctoetacff}-\ref{bctodvff} we plot the calculated $q^2$
dependence of the weak form factors of Cabibbo-Kobayashi-Maskawa (CKM)
favored  $B_c\to\eta_c$, $B_c\to J/\psi$, $B_c\to\eta_c'$, $B_c\to
\psi'$, as well as CKM suppressed  $B_c\to D$, $B_c\to D^*$
transitions in the whole kinematical range. The different behavior
(growing or falling with $q^2$) of the form factors displayed in
Figs.~\ref{bctoetacff}-\ref{bctopsiexcff} is evoked by the properties
of the final meson wave functions, since the $2S$ wave function of the
radially excited $\eta'$, $\psi'$ mesons has a zero.
  
In the following sections we use the obtained form factors for the
calculation of the $B_c$ semileptonic and nonleptonic decay rates.

\section{Semileptonic decays}\label{ssd}

The differential semileptonic decay rates can be expressed in terms of
the form factors as follows.

(a) $B_c\to Pe\nu$ decays  ($P=\eta_c,D$)
\begin{equation}
  \label{eq:dgp}
  \frac{{\rm d}\Gamma}{{\rm d}q^2}(B_c\to Pe\nu)=\frac{G_F^2 
  \Delta^3 |V_{qb}|^2}{24\pi^3} |f_+(q^2)|^2.
\end{equation}

(b) $B_c\to Ve\nu$ decays ($V=\psi,D^*$)
\begin{equation}
  \label{eq:dgv}
\frac{{\rm d}\Gamma}{{\rm d}q^2}(B_c\to Ve\nu)=\frac{G_F^2
\Delta|V_{qb}|^2}{96\pi^3}\frac{q^2}{M_{B_c}^2}
\left(|H_+(q^2)|^2+|H_-(q^2)|^2   
+|H_0(q^2)|^2\right),
\end{equation}
where $G_F$ is the Fermi constant, $V_{qb}$ is the
CKM matrix element ($q=c,u$),
\[\Delta\equiv|{\bf\Delta}|=\sqrt{\frac{(M_{B_c}^2+M_{P,V}^2-q^2)^2}
{4M_{B_c}^2}-M_{P,V}^2}.
\]
The helicity amplitudes are given by
\begin{equation}
  \label{eq:helamp}
  H_\pm(q^2)=\frac{2M_{B_c}\Delta}{M_{B_c}+M_V}\left[V(q^2)\mp
\frac{(M_{B_c}+M_V)^2}{2M_{B_c}\Delta}A_1(q^2)\right],
\end{equation}
\begin{equation}
  \label{eq:h0a}
  H_0(q^2)=\frac1{2M_V\sqrt{q^2}}\left[(M_{B_c}+M_V)
(M_{B_c}^2-M_V^2-q^2)A_1(q^2)-\frac{4M_{B_c}^2\Delta^2}{M_{B_c}
+M_V}A_2(q^2)\right].
\end{equation}
The decay rates to the longitudinally and transversely polarized
vector mesons are defined by
\begin{equation}
  \label{eq:dgl}
\frac{{\rm d}\Gamma_L}{{\rm d}q^2}=\frac{G_F^2
\Delta|V_{qb}|^2}{96\pi^3}\frac{q^2}{M_{B_c}^2}
|H_0(q^2)|^2,  
\end{equation}
\begin{equation}
  \label{eq:dgt}
\frac{{\rm d}\Gamma_T}{{\rm d}q^2}=
\frac{{\rm d}\Gamma_+}{{\rm d}q^2}+\frac{{\rm d}\Gamma_-}{{\rm d}q^2}
=\frac{G_F^2\Delta|V_{qb}|^2}{96\pi^3}\frac{q^2}{M_{B_c}^2}
\left(|H_+(q^2)|^2+|H_-(q^2)|^2\right). 
\end{equation}

\begin{figure}
  \centering
  \includegraphics{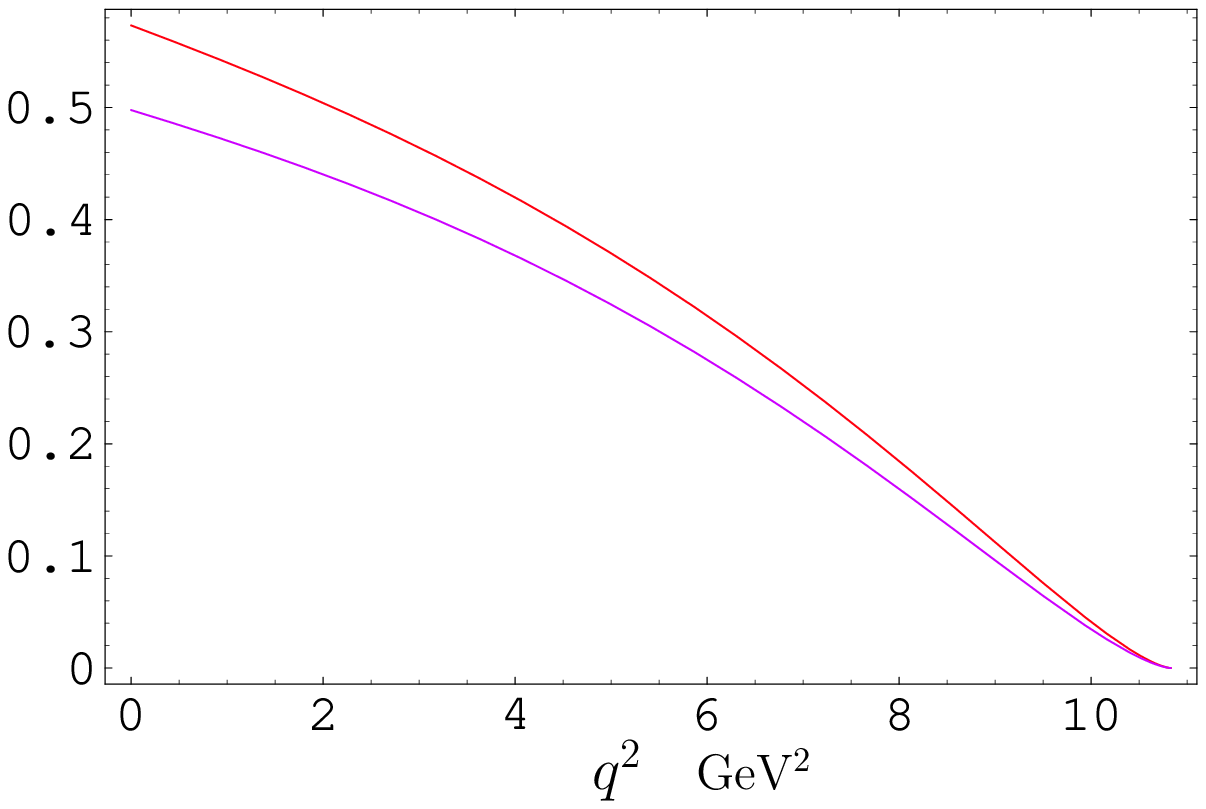}


\caption{Differential decay rates
  $(1/{|V_{cb}|^2}){{\rm d}\Gamma}/{{\rm d}q^2}$ of $B_c\to\eta_c e\nu$
  decay (in $10^{-12}$~GeV$^{-1}$). The lower curve is calculated without
  account of $1/m_{b,c}$ corrections.\label{dgbcetan}} 
\end{figure}

\begin{figure}
  \centering
  \includegraphics{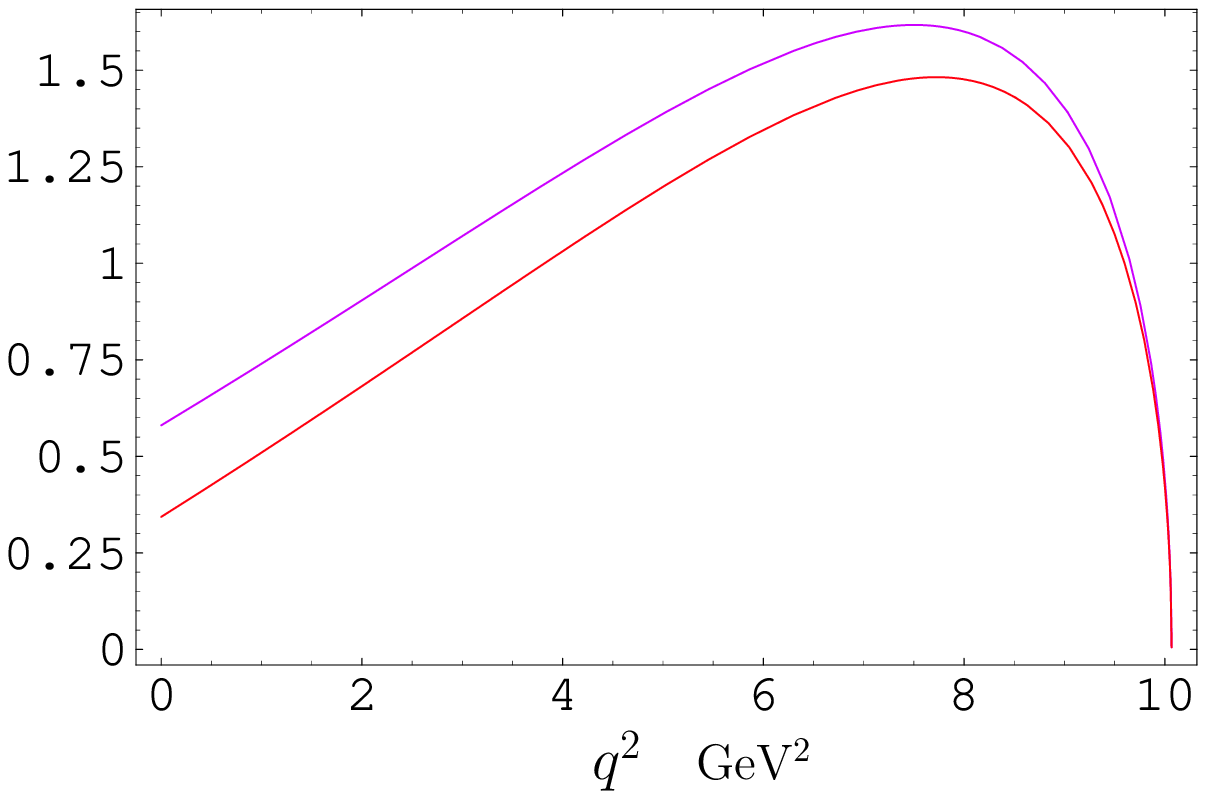}


\caption{Differential decay rates
  $(1/{|V_{cb}|^2}){{\rm d}\Gamma}/{{\rm d}q^2}$ of $B_c\to J/\psi e\nu$
  decay (in $10^{-12}$~GeV$^{-1}$). The upper curve is calculated without
  account of $1/m_{b,c}$ corrections.\label{dgbcjpsin}} 
\end{figure}

\begin{figure}
  \centering
  \includegraphics{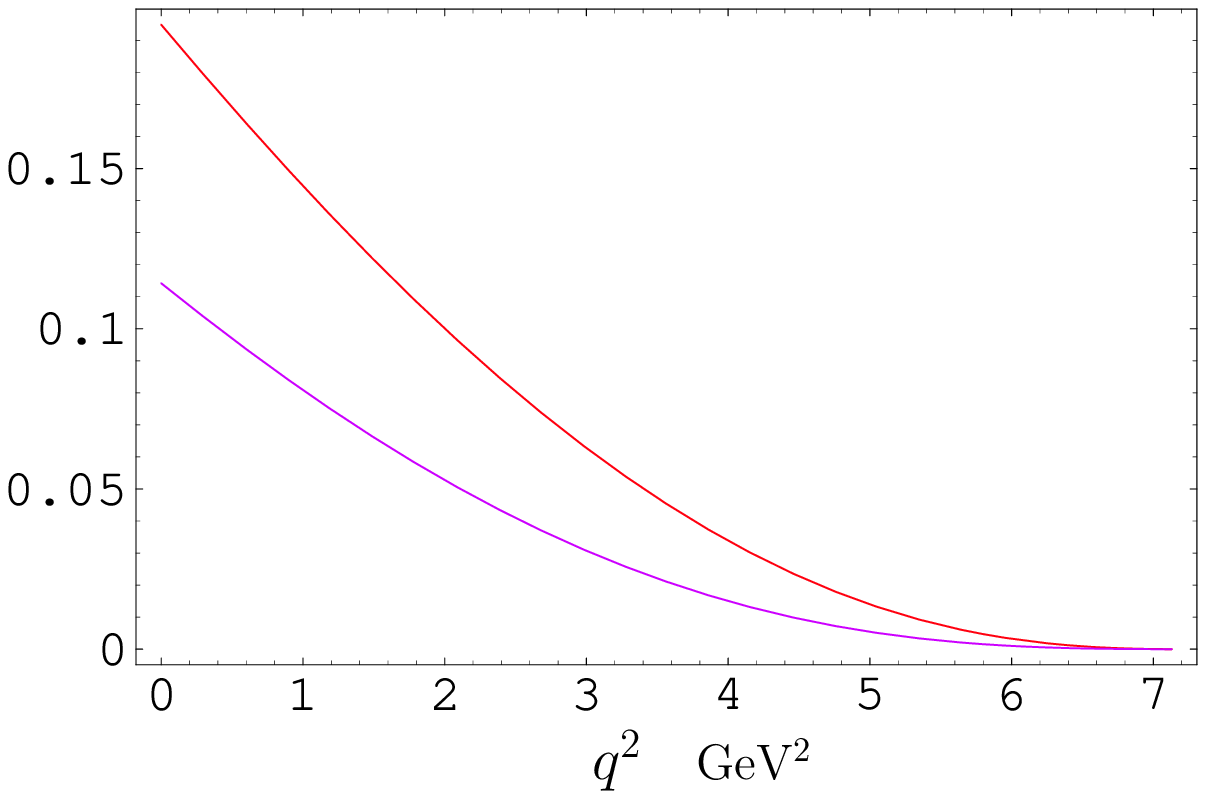} 


\caption{Differential decay rates
  $(1/{|V_{cb}|^2}){{\rm d}\Gamma}/{{\rm d}q^2}$ of $B_c\to\eta_c' e\nu$
  decay (in $10^{-12}$~GeV$^{-1}$). The lower curve is calculated without
  account of $1/m_{b,c}$ corrections.\label{dgbcetaexn}} 
\end{figure}

\begin{figure}
  \centering
  \includegraphics{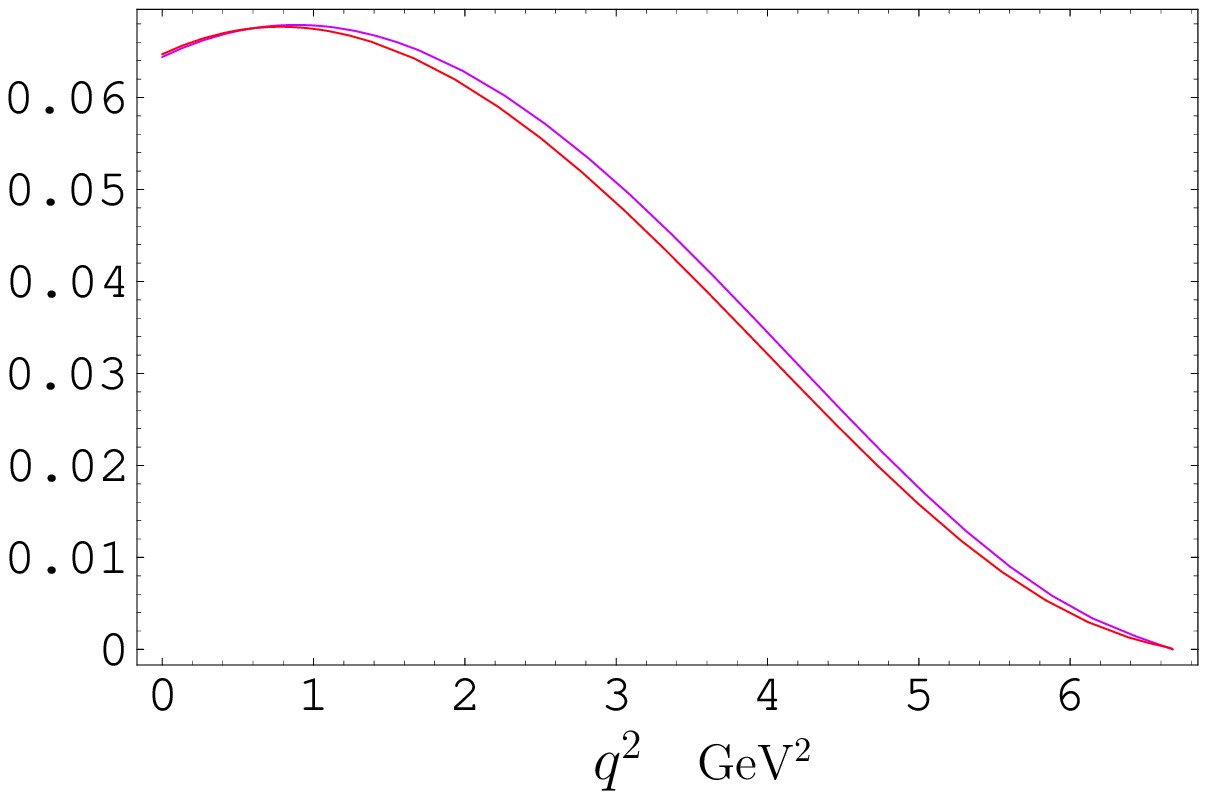}


\caption{Differential decay rates
  $(1/{|V_{cb}|^2}){{\rm d}\Gamma}/{{\rm d}q^2}$ of $B_c\to \psi' e\nu$
  decay (in $10^{-12}$~GeV$^{-1}$). The upper curve is calculated without
  account of $1/m_{b,c}$ corrections.\label{dgbcjpsiexn}}
\end{figure}

\begin{figure}
  \centering
  \includegraphics{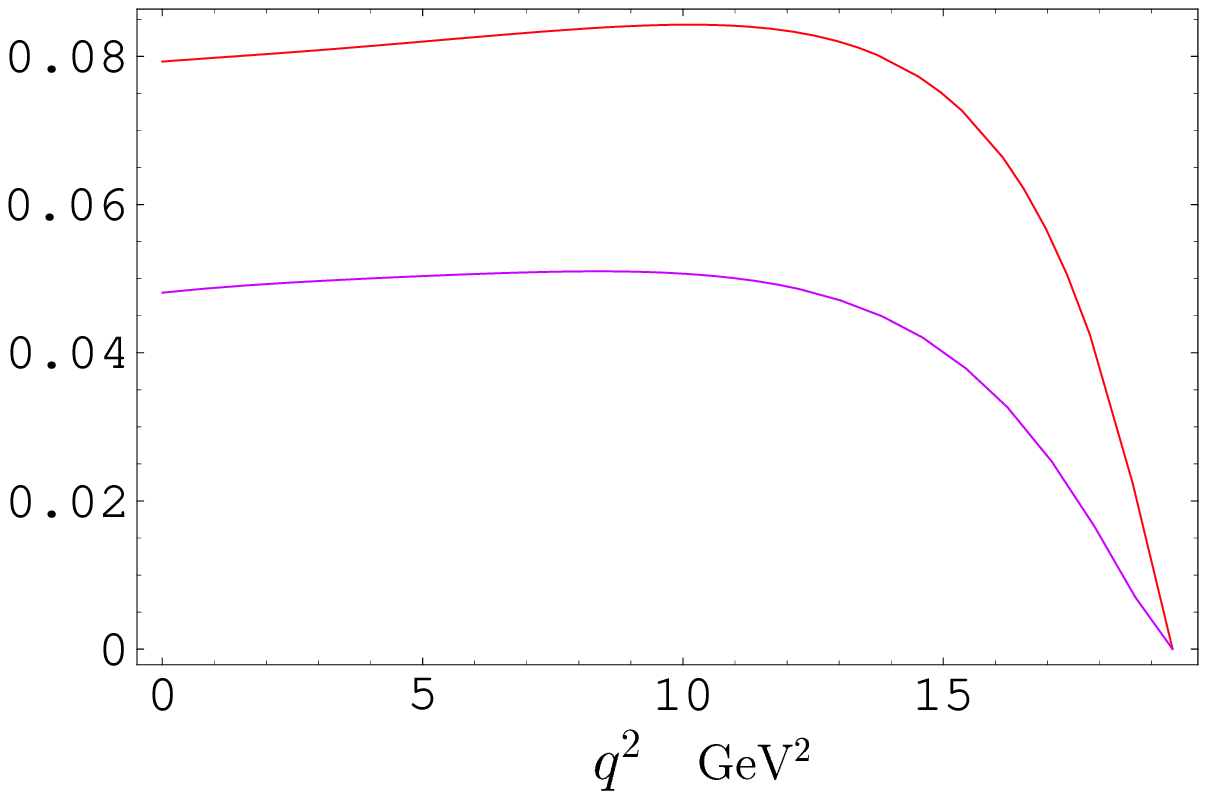}


\caption{Differential decay rate
  $(1/{|V_{ub}|^2}){{\rm d}\Gamma}/{{\rm d}q^2}$ of $B_c\to D e\nu$
  decay (in $10^{-12}$ GeV$^{-1}$). The lower curve is calculated without
  account of $1/m_{b,c}$ corrections.\label{dgbcdn}}
\end{figure}

\begin{figure}
  \centering
  \includegraphics{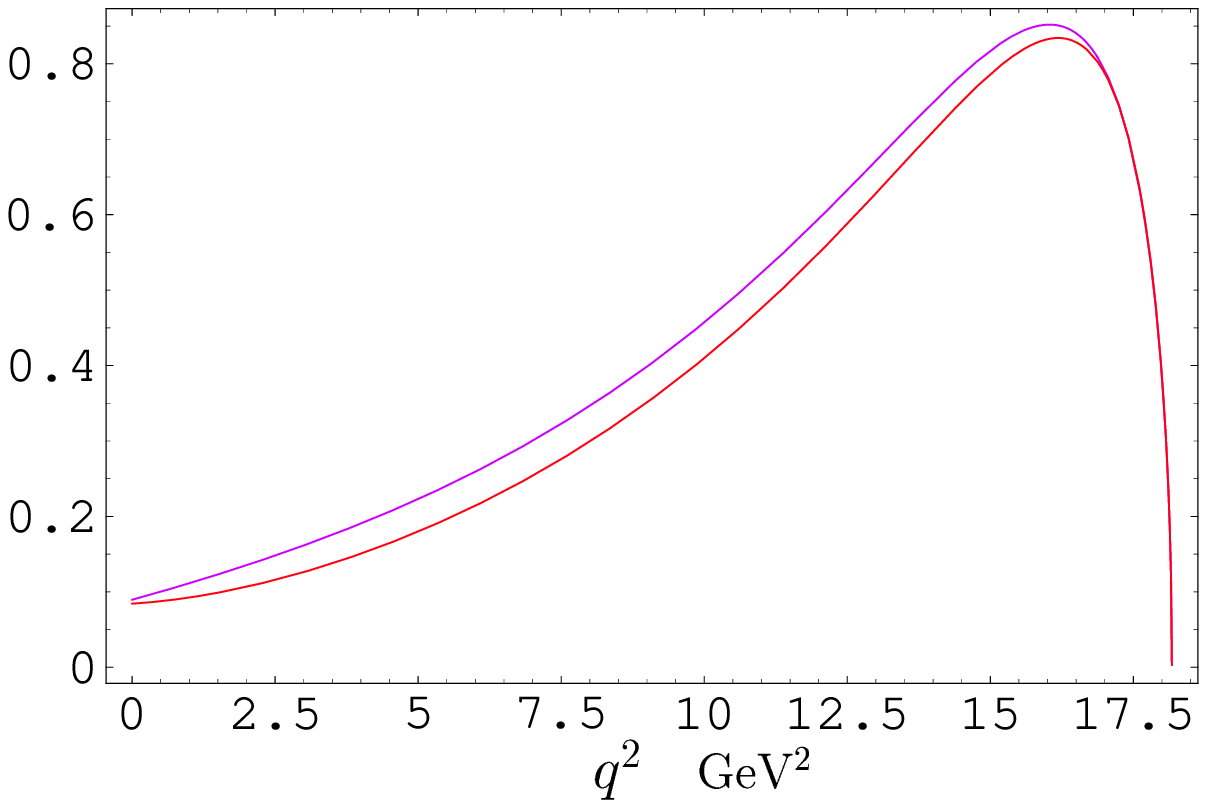}


\caption{Differential decay rates
  $(1/{|V_{ub}|^2}){{\rm d}\Gamma}/{{\rm d}q^2}$ of $B_c\to D^* e\nu$
  decay (in $10^{-12}$ GeV$^{-1}$). The upper curve is calculated without
  account of $1/m_{b,c}$ corrections.\label{dgbcdvn}}

\end{figure}

In Figs.~\ref{dgbcetan}-\ref{dgbcdvn} we plot the differential
semileptonic decay rates ${\rm d}\Gamma/{{\rm d}q^2}$ for semileptonic
decays $B_c\to \eta_c(J/\psi)e\nu$, $B_c\to \eta'_c(\psi')e\nu$ and
$B_c\to D(D^*)e\nu$  calculated in our model using
Eqs.~(\ref{eq:dgp}), (\ref{eq:dgv}) both with and without
account of $1/m_{b,c}$ corrections to the decay form factors
(\ref{eq:fpl})--(\ref{eq:a0v}).~\footnote{Relativistic wave functions
  were used for both calculations.}   From these plots we see that
relativistic effects related to heavy quarks increase semileptonic
$B_c$ decay rates to the 
pseudoscalar $\eta_c$, $\eta_c'$ and $D$ mesons, while semileptonic
decay rates to vector $J/\psi$, $\psi'$ and $D^*$ mesons are decreased
by them. The decay rates for $B_c\to \eta_c'e\nu$ and $B_c\to
De\nu$ receive the largest $1/m_{b,c}$ corrections. This is not
surprising since in the former decay the radially excited $\eta_c'$ wave
function has a zero, which considerably decreases the nonrelativistic
contribution and thus increases the relative size of relativistic
effects. In the latter decay, the role of relativistic effects is
enhanced due to the relativistic light quark in the $D$ meson.

\begin{table}
\caption{Semileptonic decay rates $\Gamma$ (in $10^{-15}$ GeV) of
  $B_c$ decays to charmonium and $D$
   mesons. } 
\label{ssdr}
\begin{ruledtabular}
\begin{tabular}{ccccccccccc}
Decay& our& \cite{iks}& \cite{klo} & \cite{emv} & \cite{cc} &
\cite{cdf} & \cite{aknt} & \cite{nw} &\cite{lyl}&\cite{lc}\\
\hline
$B_c\to\eta_c e\nu$ & 5.9 & 14 & 11 & 11.1 & 14.2 & 2.1(6.9) & 8.6 &
6.8 & 4.3& 8.31\\
$B_c\to\eta_c' e\nu$ & 0.46 &  & 0.60 &  & 0.73 & 0.3 & & &  & 0.605\\
$B_c\to J/\psi e\nu$ & 17.7 & 33 & 28 & 30.2 & 34.4 & 21.6(48.3) &
17.5 & 19.4 & 16.8 & 20.3\\
$B_c\to\psi' e\nu$ & 0.44 &  & 1.94 &  & 1.45 & 1.7 & & & & 0.186\\
$B_c\to D e\nu$ & 0.019 & 0.26 & 0.059 & 0.049 & 0.094 & 0.005(0.03) &
& & 0.001 & 0.0853\\
$B_c\to D^* e\nu$ & 0.11 & 0.49 & 0.27 & 0.192 & 0.269 & 0.12(0.5) & &
& 0.06 & 0.204
\end{tabular}
\end{ruledtabular}
\end{table}

We calculate the total rates of the semileptonic $B_c$ decays  
to the ground and radially excited states of charmonium and $D$
mesons integrating the corresponding 
differential decay rates over $q^2$. For calculations we use the
following values of the CKM matrix elements: $|V_{cb}|=0.041$,
$|V_{ub}|=0.0036$.  The results  are given in
Table~\ref{ssdr} in comparison with predictions of other
approaches based on 
quark models \cite{iks,emv,cc,aknt,nw,lc}, QCD sum
rules \cite{klo} and on the application of heavy quark
symmetry relations \cite{cdf,lyl} to the quark model. Our predictions
for the CKM favored semileptonic $B_c$ decays to charmonium ground states
are almost 2 times smaller than those of QCD sum rules \cite{klo} and
quark models \cite{iks,emv,cc}, but agree with quark model results
\cite{aknt,nw,lyl,lc}. Note that the ratios of the $B_c\to J/\psi
e\nu$ to $B_c\to \eta_c e\nu$ decay rates have close values in all
approaches except 
\cite{cdf}. In the case of semileptonic decays to radially excited
charmonium states our prediction for the decay to the pseudoscalar $\eta_c'$
state is consistent with others, while the one for the decay to
$\psi'$ is considerably smaller (with the exception of Ref.~\cite{lc}). For
the CKM suppressed 
semileptonic decays of $B_c$ to $D$ mesons our results are in agreement
with those of Ref.~\cite{cdf}.
     
In Table~\ref{sr} we present for completeness our predictions for the rates
of the semileptonic $B_c$ decays  to vector ($\psi$ and $D^*$) mesons
with longitudinal ($L$) or transverse ($T$) polarization and to the
states with helicities $\lambda=\pm 1$, as well as their ratios. 

\begin{table}
\caption{Semileptonic decay rates $\Gamma_{L,T,+,-}$ (in $10^{-15}$
  GeV) and their ratios for $B_c$ decays to vector $\psi$ and $D^*$
  mesons.} 
\label{sr}
\begin{ruledtabular}
\begin{tabular}{ccccccc}
Decay& $\Gamma_L$ & $\Gamma_T$ & $\Gamma_L/\Gamma_T$ & $\Gamma_+$ &
$\Gamma_-$ &  $\Gamma_+/\Gamma_-$\\
\hline
$B_c\to J/\psi e\nu$ & 7.8 & 9.9 & 0.78 & 2.9 & 7.0 & 0.40\\
$B_c\to\psi' e\nu$ & 0.29 & 0.15 & 1.85 & 0.05 & 0.10 & 0.47\\
$B_c\to D^* e\nu$ & 0.04 & 0.07 & 0.53 & 0.015 & 0.055 & 0.24\\
\end{tabular}
\end{ruledtabular}
\end{table}

\section{Nonleptonic decays}\label{nl}
In the standard model nonleptonic $B_c$ decays are described by the
effective Hamiltonian, obtained by integrating out the heavy $W$-boson
and top quark. For the case of $b\to c,u$ transitions, one gets
\begin{equation}
\label{heff}
H_{\rm eff}=\frac{G_F}{\sqrt{2}}V_{cb}\left[c_1(\mu)O_1^{cb}+
c_2(\mu)O_2^{cb}\right] 
+\frac{G_F}{\sqrt{2}}V_{ub}\left[c_1(\mu)O_1^{ub}+
c_2(\mu)O_2^{ub}\right] +\dots.
\end{equation}
The Wilson coefficients $c_{1,2}(\mu)$ are evaluated
perturbatively at the $W$ scale and then are evolved down to the
renormalization scale $\mu\approx m_b$ by the renormalization-group
equations. The ellipsis denote the penguin operators, the Wilson
coefficients of  which are numerically much smaller than $c_{1,2}$.
The local four-quark operators $O_1$ and $O_2$ are given by
\begin{eqnarray}
\label{o12}
O_1^{qb}&=& [({\tilde d}u)_{V-A}+({\tilde s}c)_{V-A}](\bar q
b)_{V-A}, \cr O_2^{qb}&=& (\bar qu)_{V-A}({\tilde d}b)_{V-A}+(\bar
qc)_{V-A}({\tilde s}b)_{V-A}, \qquad q=(u,c),
\end{eqnarray}
where the rotated antiquark fields are
\begin{equation} \label{ds}
\tilde d=V_{ud}\bar d+V_{us}\bar s,
 \qquad \tilde s=V_{cd}\bar d+V_{cs}\bar s,
\end{equation}
and for
the hadronic current the following notation is used
$$(\bar qq')_{V-A}=\bar q\gamma_\mu(1-\gamma_5)q' \equiv J^W_\mu.$$

The factorization approach, which is extensively used for the calculation
of two-body nonleptonic decays, such as $B_c\to FM$, assumes that the
nonleptonic decay amplitude reduces to the product of a form factor
and a decay constant \cite{bsw}. This assumption in general cannot be
exact.  However, it is expected that factorization can hold 
for the energetic decays, where the final $F$ meson is heavy and the $M$
meson is light \cite{dg}. A justification of this assumption is
usually based on the issue of color 
transparency \cite{jb}. In these decays the final hadrons
are produced in the form of point-like color-singlet objects with a
large relative momentum. And thus the hadronization of the decay
products occurs  after they are too far away for strongly interacting
with each other. That provides the possibility to avoid the final
state interaction. A more general treatment of factorization is given in
Refs.~\cite{bbns,bs}.  

\begin{figure}
  \centering 
  \includegraphics{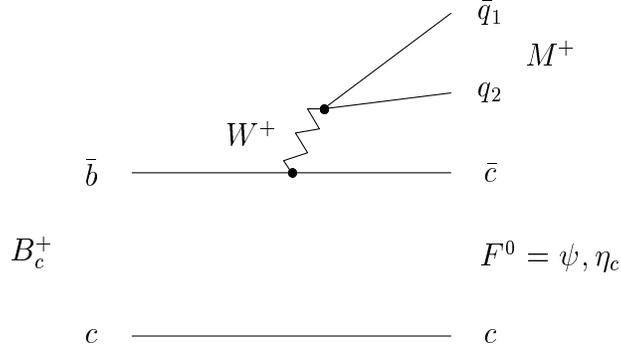}
\caption{Quark diagram for the nonleptonic $B_c^+\to F^0M^+$ decay. }
\label{d3}
\end{figure}

In this paper we limit our analysis of the $B_c^+$ nonleptonic decays to
the case when the final meson $F^0$ 
is charmonium~\footnote{We do not consider nonleptonic $B_c$ decays
  where the final meson $F$ is a $D$ meson, since such decays are
  strongly CKM suppressed and thus receive important contributions
  from the weak annihilation and penguins.} and the light $M^+$ meson is
$\pi^+$, $\rho^+$ or 
$K^{(*)+}$. The corresponding diagram is shown in
Fig.~\ref{d3}, where $q_1=d$,  $s$ and $q_2=u$. Then the decay
amplitude can be approximated by the product of one-particle matrix
elements 
\begin{equation}
\label{factor} \langle F^0M^+|H_{\rm eff}|B_c^+\rangle= \frac{G_F}{\sqrt{2}}
V_{cb}V_{q_1q_2} a_1\langle F|(\bar
bc)_{V-A}|B_c\rangle\langle M|(\bar q_1q_2)_{V-A}|0\rangle ,
\end{equation}
where
\begin{equation}
\label{amu} a_1=c_1(\mu)+\frac{1}{N_c}c_2(\mu)
\end{equation}
and $N_c$ is the number of colors. 

The matrix element of the current $J^W_\mu$ between vacuum and final
pseudoscalar ($P$) or vector ($V$) meson is parametrized by the decay
constants $f_{P,V}$
\begin{equation}
\langle P|\bar q_1 \gamma^\mu\gamma_5 q_2|0\rangle=if_Pp^\mu_P, \qquad
\langle V|\bar q_1\gamma_\mu q_2|0\rangle=\epsilon_\mu M_Vf_V.
\end{equation}
We use the following values of the decay constants: $f_\pi=0.131$~GeV,
$f_\rho=0.208$~GeV, $f_K=0.160$~GeV and $f_{K^*}=0.214$~GeV. The CKM
matrix elements are $|V_{ud}|=0.975$,  $|V_{us}|=0.222$.

\begin{table}
\caption{Nonleptonic $B_c$ decay rates $\Gamma$ (in $10^{-15}$ GeV). }
\label{nldr}
\begin{ruledtabular}
\begin{tabular}{cccccccc}
Decay& our&  \cite{klo} & \cite{emv} & \cite{cc} & \cite{aknt} &
\cite{cdf}& \cite{lc} \\
\hline
$B_c^+\to\eta_c\pi^+$ & 0.93$a_1^2$ & 1.8$a_1^2$ & 1.59$a_1^2$ &
2.07$a_1^2$ & 1.47$a_1^2$ & 0.28$a_1^2$ & 1.49$a_1^2$ \\
$B_c^+\to\eta_c\rho^+$ & 2.3$a_1^2$ & 4.5$a_1^2$ & 3.74$a_1^2$ &
5.48$a_1^2$ & 3.35$a_1^2$ & 0.75$a_1^2$ & 3.93$a_1^2$\\
$B_c^+\to J/\psi\pi^+$ & 0.67$a_1^2$ & 1.43$a_1^2$ & 1.22$a_1^2$ &
1.97$a_1^2$ & 0.82$a_1^2$ & 1.48$a_1^2$& 1.01$a_1^2$\\
$B_c^+\to J/\psi\rho^+$ & 1.8$a_1^2$ & 4.37$a_1^2$ & 3.48$a_1^2$ &
5.95$a_1^2$ & 2.32$a_1^2$ & 4.14$a_1^2$ & 3.25$a_1^2$\\
$B_c^+\to\eta_c K^+$ & 0.073$a_1^2$ & 0.15$a_1^2$ & 0.119$a_1^2$ &
0.161$a_1^2$ & 0.15$a_1^2$ & 0.023$a_1^2$ & 0.115$a_1^2$\\
$B_c^+\to\eta_c K^{*+}$ & 0.12$a_1^2$ & 0.22$a_1^2$ & 0.200$a_1^2$ &
0.286$a_1^2$ & 0.24$a_1^2$ & 0.041$a_1^2$ & 0.198$a_1^2$\\
$B_c^+\to J/\psi K^+$ & 0.052$a_1^2$ & 0.12$a_1^2$ & 0.090 $a_1^2$ &
0.152$a_1^2$ & 0.079$a_1^2$ & 0.076$a_1^2$& 0.0764$a_1^2$\\
$B_c^+\to J/\psi K^{*+}$ & 0.11$a_1^2$ & 0.25$a_1^2$ & 0.197$a_1^2$ &
0.324$a_1^2$ & 0.18$a_1^2$ &0.23$a_1^2$ & 0.174$a_1^2$\\
$B_c^+\to\eta_c'\pi^+$ & 0.19$a_1^2$ & & & 0.268$a_1^2$ & &
0.074$a_1^2$ & 0.248$a_1^2$\\ 
$B_c^+\to\eta_c'\rho^+$ & 0.40$a_1^2$ & & & 0.622$a_1^2$ & &
0.16$a_1^2$& 0.587$a_1^2$\\
$B_c^+\to \psi'\pi^+$ & 0.12$a_1^2$ & & & 0.252$a_1^2$& &
0.22$a_1^2$& 0.0708$a_1^2$\\
$B_c^+\to \psi'\rho^+$ & 0.20$a_1^2$ & & & 0.710 $a_1^2$& &
0.54$a_1^2$& 0.183$a_1^2$\\
$B_c^+\to\eta_c'K^+$ & 0.014$a_1^2$ & & & 0.020$a_1^2$ & &
0.0055$a_1^2$& 0.0184$a_1^2$\\ 
$B_c^+\to\eta_c'K^{*+}$ & 0.021$a_1^2$ & & & 0.031$a_1^2$ & &
0.008$a_1^2$& 0.0283$a_1^2$\\
$B_c^+\to \psi'K^+$ & 0.009$a_1^2$ & & & 0.018$a_1^2$& &
0.01$a_1^2$& 0.00499$a_1^2$\\
$B_c^+\to \psi'K^{*+}$ & 0.011$a_1^2$ & & & 0.038 $a_1^2$& &
0.03$a_1^2$& 0.00909$a_1^2$
\end{tabular}
\end{ruledtabular}
\end{table}

The matrix elements of the weak current between the $B_c$ meson and the final
charmonium entering the factorized nonleptonic decay
amplitude (\ref{factor}) are parametrized by the set of decay form
factors defined in Eqs.~(\ref{eq:pff1}) and (\ref{eq:vff1}). Using the
form factor values calculated in Sec.~\ref{dff}, we get predictions for
the nonleptonic $B_c^+\to F^0M^+$  decay rates and give them in
Table~\ref{nldr} in 
comparison with other calculations \cite{klo,emv,cc,aknt,cdf,lc}.
We see that for most decays our model predicts slightly lower decay
rates than other approaches.

\section{Conclusions}
\label{sec:conc}

In this paper we calculated weak semileptonic and nonleptonic $B_c$ decays
to the charmonium and $D$ meson final states. The
corresponding decay form factors were calculated in the framework of
the relativistic quark model using $v/c$ expansion for the $B_c$ meson
and charmonium and heavy quark expansion for $D$ mesons. These
transitions  proceed in a large kinematically
allowed region. As a result, the recoil momentum ${\bf \Delta}$ of the
final meson in the $B_c$ rest frame is mostly large compared to the
binding energy and the 
relative momentum of quarks forming a meson. Our approach permits  to
determine explicitly the dependence of form factors of the CKM favored
$B_c$ transition to charmonium in the whole kinematical region through the
overlap integrals of the meson wave functions. This is a real
achievement making our results more reliable, since most other
approaches determine form 
factors only at the single point of zero ($q^2=q^2_{\rm max}$) or
maximum ($q^2=0$) recoil of the final meson and then the extrapolation
is used.  We
calculated form factors of weak $B_c$ transitions 
to charmonium up to second order corrections in the ratio of the
relative quark momentum to the heavy ($b$ and $c$) quark mass.    
In the case of the CKM suppressed $B_c$ transitions to $D$ mesons the
situation is more complicated, since the active quark undergoes
heavy-to-light transition. The leading contribution as in the previous
case can be determined exactly, while the subleading contribution,
which is suppressed by the small binding energy, can be determined
reliably  in most part of the kinematical range except a small
region  near the point of zero recoil ($q^2=q^2_{\rm max}$) of the final
$D$ meson. As the numerical analysis shows, the extrapolation of form
factors obtained in such a way to the region of small recoil introduces
only minor uncertainties.

\begin{table}
\caption{Branching fractions (in \%) of exclusive $B_c$ decays calculated
  for the fixed values of the $B_c$ lifetime $\tau_{B_c}=0.46$~ps and
  $a_1=1.14$.}
\label{Br}
\begin{ruledtabular}
\begin{tabular}{cccccc}
Decay& Br & Decay & Br & Decay & Br\\
\hline
$B_c\to \eta_c e\nu$& 0.42 &$B_c^+\to\eta_c\pi^+ $& 0.085  &
$B_c^+\to \eta' \pi^{+}$& 0.017 \\ 
$B_c\to \eta'_c e\nu$& 0.032 &$B_c^+\to\eta_c\rho^+ $& 0.21 & $B_c^+\to
\eta_c' \rho^{+}$& 0.036 \\
$B_c\to J/\psi e\nu$& 1.23 &$B_c^+\to J/\psi\pi^+ $& 0.061  &$B_c^+\to
\psi' \pi^{+}$& 0.011\\
 $B_c\to \psi' e\nu$& 0.031 &$B_c^+\to J/\psi \rho^+$& 0.16 &
 $B_c^+\to \psi' \rho^{+}$& 0.018 \\ 
$B_c\to D e\nu$& 0.013 & $B_c^+\to \eta_c K^+$& 0.007 &$B_c^+\to
\eta_c' K^+$& 0.001\\
$B_c\to D^* e\nu$& 0.037 &$B_c^+\to \eta_c K^{*+}$& 0.011&$B_c^+\to
\eta_c' K^{*+}$& 0.002\\
& & $B_c^+\to J/\psi K^{+}$& 0.005 & $B_c^+\to
\psi' K^{+}$& 0.001\\
& &$B_c^+\to J/\psi K^{*+}$& 0.010  &$B_c^+\to
\psi' K^{*+}$& 0.001\\
\end{tabular}
\end{ruledtabular}
\end{table}

We calculated semileptonic and nonleptonic (in factorization
approximation) $B_c$ decay rates. 
Our predictions for the branching fractions are summarized in
Table~\ref {Br}, where we use the central experimental value of the
$B_c$ meson lifetime \cite{pdg}. From this table we see that
the considered semileptonic 
decays to charmonium and $D$ mesons give in total 1.72\% of the $B_c$
decay rate, while the energetic nonleptonic decays give
additional 0.63\%. It is expected that the dominant contribution to
the $B_c$ total rate comes from the charmed quark decays. These decays
will be considered in a forthcoming publication.

\acknowledgments
The authors express their gratitude to M. M\"uller-Preussker and
V. Savrin for support and discussions.
Two of us (R.N.F and V.O.G.) were supported in part by the 
{\it Deutsche Forschungsgemeinschaft} under contract Eb 139/2-2.


\appendix*
\section{Form factors of weak $\bm{B_{\lowercase{c}}}$ decays}

(a) $B_c\to P$ transition ($P=\eta_c,D$)

\begin{eqnarray}
  \label{eq:fpl}
  f_+^{(1)}(q^2)&=&\sqrt{\frac{E_P}{M_{B_c}}}\int \frac{{\rm
  d}^3p}{(2\pi)^3} \bar\Psi_P\left({\bf p}+\frac{2m_c}{E_P+M_P}{\bf
  \Delta} \right)\sqrt{\frac{\epsilon_q(p+\Delta)+
  m_q}{2\epsilon_q(p+\Delta)}} \sqrt{\frac{\epsilon_b(p)+
  m_b}{2\epsilon_b(p)}}\cr
&&\times\Biggl\{1+\frac{M_{B_c}-E_P}{\epsilon_q(p+\Delta)+m_q}+\frac{({\bf
p\Delta})}{{\bf \Delta}^2}\Bigglb(\frac{{\bf
\Delta}^2}{[\epsilon_q(p+\Delta)+m_q][\epsilon_b(p)+m_b]}+(M_{B_c}-E_P) \cr
&&\times\left(\frac1{\epsilon_q(p+\Delta)+m_q}+
\frac1{\epsilon_b(p)+m_b}\right) \Biggrb)+\frac{{\bf
  p}^2}{[\epsilon_q(p+\Delta)+m_q][\epsilon_b(p)+m_b]} \cr
&&+\frac23{\bf p}^2\Bigglb(\frac{E_P-M_P}
{[\epsilon_q(p+\Delta)+m_q][\epsilon_b(p)+m_b]}
\left(\frac1{\epsilon_q(p+\Delta)+m_q}-\frac1{\epsilon_c(p)+m_c}\right)\cr
&&+\frac{M_{B_c}-E_P}{E_P+M_P}\left(\frac1{\epsilon_q(p+\Delta)+m_q}-
\frac1{\epsilon_b(p)+m_b}\right)\cr
&&\times\left(\frac1{\epsilon_c(p)+m_c}-
\frac1{\epsilon_q(p+\Delta)+m_q}\right)\Biggrb)\Biggr\}\Psi_{B_c}({\bf p}),  
\end{eqnarray}

\begin{eqnarray}
  \label{eq:fpls}
  f_+^{S(2)}(q^2)&=&\sqrt{\frac{E_P}{M_{B_c}}}\int \frac{{\rm
  d}^3p}{(2\pi)^3} \bar\Psi_P\left({\bf p}+\frac{2m_c}{E_P+M_P}{\bf
  \Delta} \right)\sqrt{\frac{\epsilon_q(\Delta)+
  m_q}{2\epsilon_q(\Delta)}} \Biggl\{
  \frac1{\epsilon_q(\Delta)}\Biggl( \frac{\epsilon_q(\Delta)-m_q}
  {\epsilon_q(\Delta)+m_q} \cr
&& -\frac{M_{B_c}-E_P}{\epsilon_q(\Delta)+mq}\Biggr)
\left[M_P-\epsilon_q\left(p+\frac{2m_c}{E_P+M_P}\Delta\right)
-\epsilon_c\left(p+\frac{2m_c}{E_P+M_P}\Delta \right)\right]
  \cr
&&+\frac{({\bf p\Delta})}{{\bf \Delta}^2}\Bigglb(
\Biggl[\frac1{2\epsilon_q(\Delta)}\left(\frac{\epsilon_q(\Delta)-m_q}
{\epsilon_q(\Delta)+m_q}
  -\frac{M_{B_c}-E_P}{\epsilon_q(\Delta)+m_q}\right)
  -\frac1{2m_b}\Biggl(\frac{\epsilon_q(\Delta)-m_q} 
{\epsilon_b(\Delta)+m_b}\cr
&&+\frac{M_{B_c}-E_P}{\epsilon_b(\Delta)+m_b}\Biggr)\Biggr] 
\Biggl[M_{B_c}+M_P-\epsilon_b(p)-\epsilon_c(p)
-\epsilon_q\left(p+\frac{2m_c}{E_P+M_P}\Delta\right)\cr
&&-\epsilon_c\left(p+\frac{2m_c}{E_P+M_P}\Delta \right)\Biggr]
-\frac{\epsilon_q(\Delta)-m_q}{2m_b\epsilon_q(\Delta)}
  \left(1-\frac{M_{B_c}-E_P}{\epsilon_q(\Delta)+m_q}\right) \cr
&&\times\left[M_P-\epsilon_q\left(p+\frac{2m_c}{E_P+M_P}\Delta\right)
-\epsilon_c\left(p+\frac{2m_c}{E_P+M_P}\Delta \right)\right]
\Biggrb)\Biggr\}\Psi_{B_c}({\bf p}),
\end{eqnarray}

\begin{eqnarray}
  \label{eq:fplv}
  f_+^{V(2)}(q^2)&=&\sqrt{\frac{E_P}{M_{B_c}}}\int \frac{{\rm
  d}^3p}{(2\pi)^3} \bar\Psi_P\left({\bf p}+\frac{2m_c}{E_P+M_P}{\bf
  \Delta} \right)\sqrt{\frac{\epsilon_q(\Delta)+
  m_q}{2\epsilon_q(\Delta)}}\frac{({\bf p\Delta})}{{\bf
  \Delta}^2}\frac1{2m_c} \cr
&&\times\Biggl\{ 
\Biggl[(\epsilon_q(\Delta)-m_q)\left(\frac{1}
{\epsilon_b(\Delta)+m_b}
  -\frac{1}{\epsilon_q(\Delta)+m_q}\right)
  +(M_{B_c}-E_P)\Biggl(\frac{1} 
{\epsilon_b(\Delta)+m_b}\cr
&&
+\frac{1}{\epsilon_q(\Delta)+m_q}\Biggr)\Biggr] 
\Biggl[M_{B_c}+M_P-\epsilon_b(p)-\epsilon_c(p)
-\epsilon_q\left(p+\frac{2m_c}{E_P+M_P}\Delta\right)\cr
&&-\epsilon_c\left(p+\frac{2m_c}{E_P+M_P}\Delta \right)\Biggr]
+2\frac{\epsilon_q(\Delta)-m_q}{\epsilon_q(\Delta)}
  \left(\frac{\epsilon_q(\Delta)-m_q}
{\epsilon_q(\Delta)+m_q}
  -\frac{M_{B_c}-E_P}{\epsilon_q(\Delta)+m_q}\right) \cr
&&\times\left[M_{B_c}-\epsilon_b(p)-\epsilon_c(p)\right]
-\frac{2m_c(\epsilon_q(\Delta)-m_q)}{\epsilon_q(\Delta)
[\epsilon_q(\Delta)+m_q]} \Biggl[M_P
-\epsilon_q\left(p+\frac{2m_c}{E_P+M_P}\Delta\right)\cr
&&-\epsilon_c\left(p+\frac{2m_c}{E_P+M_P}\Delta \right)\Biggr]
\Biggr\}\Psi_{B_c}({\bf p}),
\end{eqnarray}

\begin{eqnarray}
  \label{eq:f01}
  f_0^{(1)}(q^2)&=&\frac{2\sqrt{E_PM_{B_c}}}{M_{B_c}^2-M_P^2}\int \frac{{\rm
  d}^3p}{(2\pi)^3} \bar\Psi_P\left({\bf p}+\frac{2m_c}{E_P+M_P}{\bf
  \Delta} \right)\sqrt{\frac{\epsilon_q(p+\Delta)+
  m_q}{2\epsilon_q(p+\Delta)}} \sqrt{\frac{\epsilon_b(p)+
  m_b}{2\epsilon_b(p)}}\cr
&&\times\Biggl\{(E_P+M_P)\left(\frac{E_P-M_P}{\epsilon_q(p+\Delta)+m_q}+
\frac{M_{B_c}-E_P}{E_P+M_P}\right)+({\bf
p\Delta})\Bigglb(\frac{1}{\epsilon_q(p+\Delta)+m_q}\cr
&&+\frac{1}{\epsilon_b(p)+m_b}+\frac{M_{B_c}-E_P}
{[\epsilon_q(p+\Delta)+m_q][\epsilon_b(p)+m_b]} \Biggrb)+\frac{{\bf
  p}^2(M_{B_c}-E_P)}{[\epsilon_q(p+\Delta)+m_q][\epsilon_b(p)+m_b]} \cr
&&-\frac23{\bf p}^2(E_P-M_P)
\left(\frac1{\epsilon_q(p+\Delta)+m_q}-\frac1{\epsilon_c(p)+m_c}\right)
\Biggl(\frac1{\epsilon_q(p+\Delta)+m_q}\cr
&&-
\frac1{\epsilon_b(p)+m_b}-\frac{M_{B_c}-E_P}
{[\epsilon_q(p+\Delta)+m_q][\epsilon_b(p)+m_b]}\Biggr)
\Biggr\}\Psi_{B_c}({\bf p}),  
\end{eqnarray}

\begin{eqnarray}
\label{eq:f0s}
f_0^{S(2)}(q^2)&=&\frac{2\sqrt{E_PM_{B_c}}}{M_{B_c}^2-M_P^2}\int \frac{{\rm
d}^3p}{(2\pi)^3} \bar\Psi_P\left({\bf p}+\frac{2m_c}{E_P+M_P}{\bf
\Delta} \right)\sqrt{\frac{\epsilon_q(\Delta)+
m_q}{2\epsilon_q(\Delta)}}(\epsilon_q(\Delta)-m_q)\Biggl\{
  \frac1{\epsilon_q(\Delta)}\cr
&&\times\Biggl(\frac{M_{B_c}-E_P}{\epsilon_q(\Delta)+mq}-1\Biggr)
\left[M_P-\epsilon_q\left(p+\frac{2m_c}{E_P+M_P}\Delta\right)
-\epsilon_c\left(p+\frac{2m_c}{E_P+M_P}\Delta \right)\right]
  \cr
&&-\frac{({\bf p\Delta})}{{\bf \Delta}^2}\Bigglb(
\Biggl[\frac{\epsilon_q(\Delta)+m_q}{2m_b[\epsilon_b(\Delta)+m_b]}
\Biggl(1 +\frac{M_{B_c}-E_P}{\epsilon_q(\Delta)+m_q}\Biggr)
-\frac1{2\epsilon_q(\Delta)}\Biggl(\frac{M_{B_c}-E_P}
{\epsilon_q(\Delta)+m_q}-1\Biggr)\Biggr]\cr 
&&\times\Biggl[M_{B_c}+M_P-\epsilon_b(p)-\epsilon_c(p)
-\epsilon_q\left(p+\frac{2m_c}{E_P+M_P}\Delta\right)
-\epsilon_c\left(p+\frac{2m_c}{E_P+M_P}\Delta \right)\Biggr]\cr
&&+\frac{\epsilon_q(\Delta)+m_q}{2m_b\epsilon_q(\Delta)}
  \left(\frac{M_{B_c}-E_P}{\epsilon_q(\Delta)+m_q}- 
\frac{\epsilon_q(\Delta)-m_q}{\epsilon_q(\Delta)+m_q}\right) 
\Biggl[M_P-\epsilon_q\left(p+\frac{2m_c}{E_P+M_P}\Delta\right)
\cr &&
-\epsilon_c\left(p+\frac{2m_c}{E_P+M_P}\Delta \right)\Biggr]
\Biggrb)\Biggr\}\Psi_{B_c}({\bf p}), 
\end{eqnarray}

\begin{eqnarray}
\label{eq:f0v}
f_0^{V(2)}(q^2)&=&\frac{2\sqrt{E_PM_{B_c}}}{M_{B_c}^2-M_P^2}\int \frac{{\rm
d}^3p}{(2\pi)^3} \bar\Psi_P\left({\bf p}+\frac{2m_c}{E_P+M_P}{\bf
\Delta} \right)\sqrt{\frac{\epsilon_q(\Delta)+
m_q}{2\epsilon_q(\Delta)}}\ \frac{({\bf p\Delta})}{2m_c}\cr
&&\times\Biggl\{ 
\Biggl[\frac{1}{\epsilon_b(\Delta)+m_b}
\Biggl(1 +\frac{M_{B_c}-E_P}{\epsilon_q(\Delta)+m_q}\Biggr)
  -\frac{m_c} 
{\epsilon_c(\Delta)[\epsilon_c(\Delta)+m_c]}\Biggl(\frac{M_{B_c}-E_P}
{\epsilon_q(\Delta)+m_q}-1\Biggr)\Biggr] \cr
&&
\times\Biggl[M_{B_c}+M_P-\epsilon_b(p)-\epsilon_c(p)
-\epsilon_q\left(p+\frac{2m_c}{E_P+M_P}\Delta\right)
-\epsilon_c\left(p+\frac{2m_c}{E_P+M_P}\Delta \right)\Biggr]\cr
&&
+\frac{\epsilon_q(\Delta)-m_q}{\epsilon_q(\Delta)[\epsilon_q(\Delta)+m_q]}
\left[M_{B_c}-\epsilon_b(p)-\epsilon_c(p)\right]
-\frac{1}{\epsilon_q(\Delta)}\left(\frac{M_{B_c}-E_P}{\epsilon_q(\Delta)+m_q}- 
\frac{\epsilon_q(\Delta)-m_q}{\epsilon_q(\Delta)+m_q}\right)\cr
&&\times \Biggl[M_P
-\epsilon_q\left(p+\frac{2m_c}{E_P+M_P}\Delta\right)
-\epsilon_c\left(p+\frac{2m_c}{E_P+M_P}\Delta \right)\Biggr]
\Biggr\}\Psi_{B_c}({\bf p}),
\end{eqnarray}
where 
\[ \left|{\bf \Delta}\right|=\sqrt{\frac{(M_{B_c}^2+M_P^2-q^2)^2}
{4M_{B_c}^2}-M_P^2},\]
\[ E_P=\sqrt{M_P^2+{\bf \Delta}^2}, \quad
 \epsilon_Q(p+\lambda
\Delta)=\sqrt{m_Q^2+({\bf p}+\lambda{\bf \Delta})^2} \quad (Q=b,c,u), \]
and the subscript $q$ corresponds to $c$ or $u$ quark for the final
$\eta_c$ or $D$ meson, respectively. 

\bigskip
(b) $B_c\to V$ transition ($V=\psi,D^*$)

\begin{eqnarray}
  \label{eq:v1}
  V^{(1)}(q^2)&=&\frac{M_{B_c}+M_V}{2\sqrt{M_{B_c}M_V}}\int \frac{{\rm
  d}^3p}{(2\pi)^3} \bar\Psi_V\left({\bf p}+\frac{2m_c}{E_V+M_V}{\bf
  \Delta} \right)\sqrt{\frac{\epsilon_q(p+\Delta)+
  m_q}{2\epsilon_q(p+\Delta)}} \sqrt{\frac{\epsilon_b(p)+
  m_b}{2\epsilon_b(p)}}\cr
&&\times\frac{2\sqrt{E_VM_V}}{\epsilon_q(p+\Delta)+m_q}\Biggl\{1
+\frac{({\bf p\Delta})}{{\bf\Delta}^2}\left(1-\frac{\epsilon_q(p+
\Delta)+m_q}{2m_b}\right)+\frac23\frac{{\bf p}^2}{E_V+M_V}\cr
&&\times\Biggl(
\frac{\epsilon_q(p+\Delta)+m_q}{2m_b[\epsilon_c(p)+m_c]}
-\frac1{\epsilon_q(p+\Delta)+m_q}\Biggr)\Biggr\}\Psi_{B_c}({\bf p}),
\end{eqnarray}

\begin{eqnarray}
  \label{eq:vs}
  V^{S(2)}(q^2)&=&\frac{M_{B_c}+M_V}{2\sqrt{M_{B_c}M_V}}\int \frac{{\rm
  d}^3p}{(2\pi)^3} \bar\Psi_V\left({\bf p}+\frac{2m_c}{E_V+M_V}{\bf
  \Delta} \right)\sqrt{\frac{\epsilon_q(\Delta)+
  m_q}{2\epsilon_q(\Delta)}}\frac{2\sqrt{E_VM_V}}
{\epsilon_q(\Delta)+m_q}\cr
&&\times\Biggl\{-\frac1{\epsilon_q(\Delta)}
\left[M_V-\epsilon_q\left(p+\frac{2m_c}{E_V+M_V}\Delta\right)
-\epsilon_c\left(p+\frac{2m_c}{E_V+M_V}\Delta \right)\right]
 \cr
&&-\frac{({\bf p\Delta})}{{\bf \Delta}^2}\frac12
\Bigglb(\left(\frac{1}{\epsilon_q(\Delta)}
-\frac{\epsilon_q(\Delta)+m_q}{m_b[\epsilon_b(\Delta)+m_b]}\right)
\Biggl[M_{B_c}+M_V-\epsilon_b(p)-\epsilon_c(p)\cr
&&-\epsilon_q\left(p+\frac{2m_c}{E_V+M_V}\Delta\right)
-\epsilon_c\left(p+\frac{2m_c}{E_V+M_V}\Delta \right)\Biggr]
+\frac{\epsilon_q(\Delta)-m_q}{2m_b\epsilon_q(\Delta)}\cr 
&&\times  
\Biggl[M_V-\epsilon_q\left(p+\frac{2m_c}{E_V+M_V}\Delta\right)
-\epsilon_c\left(p+\frac{2m_c}{E_V+M_V}\Delta \right)\Biggr]
\Biggrb) \Biggr\}\Psi_{B_c}({\bf p}), 
\end{eqnarray}

\begin{eqnarray}
  \label{eq:vv}
V^{V(2)}(q^2)&=&\frac{M_{B_c}+M_V}{2\sqrt{M_{B_c}M_V}}\int \frac{{\rm
d}^3p}{(2\pi)^3} \bar\Psi_V\left({\bf p}+\frac{2m_c}{E_V+M_V}{\bf
\Delta} \right)\sqrt{\frac{\epsilon_q(\Delta)+
m_q}{2\epsilon_q(\Delta)}}\frac{2\sqrt{E_VM_V}}
{\epsilon_q(\Delta)+m_q}\cr
&&\times\frac{({\bf p\Delta})}{{\bf \Delta}^2}\frac1{2m_c}
\left(\frac{m_q}{\epsilon_q(\Delta)}
-\frac{\epsilon_q(\Delta)+m_q}{\epsilon_b(\Delta)+m_b}\right)
\Biggl[M_{B_c}+M_V-\epsilon_b(p)-\epsilon_c(p)\cr
&&-\epsilon_q\left(p+\frac{2m_c}{E_V+M_V}\Delta\right)
-\epsilon_c\left(p+\frac{2m_c}{E_V+M_V}\Delta \right)\Biggr]
\Psi_{B_c}({\bf p}), 
\end{eqnarray}

\begin{eqnarray}
  \label{eq:a11}
  A_1^{(1)}(q^2)&=&\frac{2\sqrt{M_{B_c}M_V}}{M_{B_c}+M_V}
\sqrt{\frac{E_V}{M_V}}\int \frac{{\rm
  d}^3p}{(2\pi)^3} \bar\Psi_V\left({\bf p}+\frac{2m_c}{E_V+M_V}{\bf
  \Delta} \right)\sqrt{\frac{\epsilon_q(p+\Delta)+
  m_q}{2\epsilon_q(p+\Delta)}} \sqrt{\frac{\epsilon_b(p)+
  m_b}{2\epsilon_b(p)}}\cr
&&\times\Biggl\{1+\frac{1}{2m_b[\epsilon_q(p+\Delta)+m_q]}
\left[\frac23{\bf p}^2\frac{E_V-M_V}{\epsilon_c(p)+m_c}-
\frac{{\bf p}^2}3-
({\bf p\Delta})\right]\Biggr\}\Psi_{B_c}({\bf p}),
\end{eqnarray}

\begin{eqnarray}
  \label{eq:a1s}
  A_1^{S(2)}(q^2)&=&\frac{2\sqrt{M_{B_c}M_V}}{M_{B_c}+M_V}
\sqrt{\frac{E_V}{M_V}}\int \frac{{\rm
d}^3p}{(2\pi)^3} \bar\Psi_V\left({\bf p}+\frac{2m_c}{E_V+M_V}{\bf
\Delta} \right)\sqrt{\frac{\epsilon_q(\Delta)+m_q}{2\epsilon_q(\Delta)}}
\cr &&\times\Biggl\{\frac{\epsilon_q(\Delta)-m_q}
{\epsilon_q(\Delta)[\epsilon_q(\Delta)+m_q]}
\left[M_V-\epsilon_q\left(p+\frac{2m_c}{E_V+M_V}\Delta\right)
-\epsilon_c\left(p+\frac{2m_c}{E_V+M_V}\Delta \right)\right]\cr
&& +\frac{({\bf p\Delta})}{{\bf \Delta}^2}\frac{\epsilon_q(\Delta)-m_q}2
\Bigglb(\left(\frac{1}{\epsilon_q(\Delta)[\epsilon_q(\Delta)+m_q]}
+\frac{1}{m_b[\epsilon_b(\Delta)+m_b]}\right)
\Biggl[M_{B_c}+M_V\cr
&&-\epsilon_b(p)-\epsilon_c(p)
-\epsilon_q\left(p+\frac{2m_c}{E_V+M_V}\Delta\right)
-\epsilon_c\left(p+\frac{2m_c}{E_V+M_V}\Delta \right)\Biggr]
+\frac{1}{m_b\epsilon_q(\Delta)}\cr 
&&\times  
\Biggl[M_V-\epsilon_q\left(p+\frac{2m_c}{E_V+M_V}\Delta\right)
-\epsilon_c\left(p+\frac{2m_c}{E_V+M_V}\Delta \right)\Biggr]
\Biggrb)\Biggr\}\Psi_{B_c}({\bf p}), 
\end{eqnarray}

\begin{eqnarray}
  \label{eq:a1v}
  A_1^{V(2)}(q^2)&=&\frac{2\sqrt{M_{B_c}M_V}}{M_{B_c}+M_V}
\sqrt{\frac{E_V}{M_V}}\int \frac{{\rm
d}^3p}{(2\pi)^3} \bar\Psi_V\left({\bf p}+\frac{2m_c}{E_V+M_V}{\bf
\Delta} \right)\sqrt{\frac{\epsilon_q(\Delta)+m_q}{2\epsilon_q(\Delta)}}
\cr 
&&\times
\frac{({\bf p\Delta})}{{\bf \Delta}^2}\frac{\epsilon_q(\Delta)-m_q}{2m_c}
\Biggl\{\frac{1}{\epsilon_q(\Delta)}\Bigglb(
\frac{\epsilon_q(\Delta)-m_q}{\epsilon_q(\Delta)+m_q}[M_{B_c}-
\epsilon_b(p)-\epsilon_c(p)]\cr
&&-\Biggl[M_V-\epsilon_q\left(p+\frac{2m_c}{E_V+M_V}\Delta\right)
-\epsilon_c\left(p+\frac{2m_c}{E_V+M_V}\Delta \right)\Biggr]\Biggrb)
\cr &&
-\left(\frac{1}{\epsilon_b(\Delta)+m_b}
+\frac{m_q}{\epsilon_q(\Delta)[\epsilon_q(\Delta)+m_q]}\right)
\Biggl[M_{B_c}+M_V-\epsilon_b(p)-\epsilon_c(p)\cr
&&
-\epsilon_q\left(p+\frac{2m_c}{E_V+M_V}\Delta\right)
-\epsilon_c\left(p+\frac{2m_c}{E_V+M_V}\Delta \right)\Biggr]
\Biggr\}\Psi_{B_c}({\bf p}), 
\end{eqnarray}

\begin{eqnarray}
  \label{eq:a21}
  A_2^{(1)}(q^2)&=&\frac{M_{B_c}+M_V}{2\sqrt{M_{B_c}M_V}}
\frac{2\sqrt{E_VM_V}}{E_V+M_V}\int \frac{{\rm
  d}^3p}{(2\pi)^3} \bar\Psi_V\left({\bf p}+\frac{2m_c}{E_V+M_V}{\bf
  \Delta} \right)\sqrt{\frac{\epsilon_q(p+\Delta)+
  m_q}{2\epsilon_q(p+\Delta)}}\cr
&&\times \sqrt{\frac{\epsilon_b(p)+ m_b}{2\epsilon_b(p)}}
\Biggl\{1+\frac{M_V}{M_{B_c}}\left(1-\frac{E_V+M_V}{\epsilon_q(p+
\Delta)+m_q}\right)
-\frac{({\bf p\Delta})}{{\bf\Delta}^2}\frac{E_V+M_V}{\epsilon_q(p+
\Delta)+m_q}\cr
&&\times\Bigglb(\frac{E_V+M_V}{2m_b}
\left[1-\frac{M_V}{M_{B_c}}\left(1-\frac{\epsilon_q(p+
\Delta)+m_q}{E_V+M_V}\right)\right]+\frac{M_V}{M_{B_c}}\Biggrb)
\cr
&&+\frac23\frac{{\bf p}^2}{\epsilon_q(p+\Delta)+m_q}
\Bigglb(\frac1{2m_b}\Biggl[\frac{E_V+M_V}{\epsilon_c(p)+m_c}
-\frac12+\frac{M_V}{\epsilon_q(p+\Delta)+m_q}\cr
&&+
\frac{M_V}{M_{B_c}}\left(
\frac{\epsilon_q(p+\Delta)+m_q}{\epsilon_c(p)+m_c}-\frac{E_V+M_V}
{\epsilon_c(p)+m_c}+\frac12
-\frac{E_V}{\epsilon_q(p+\Delta)+m_q}\right)\Biggr]\cr
&&+\frac{M_V}{M_{B_c}}\left(\frac1{\epsilon_q(p+\Delta)+m_q}
+\frac1{\epsilon_c(p)+m_c}\right)\Biggrb)\Biggr\}\Psi_{B_c}({\bf p}),
\end{eqnarray}

\begin{eqnarray}
  \label{eq:a2s}
  A_2^{S(2)}(q^2)&=&\frac{M_{B_c}+M_V}{2\sqrt{M_{B_c}M_V}}
\frac{2\sqrt{E_VM_V}}{E_V+M_V}\int \frac{{\rm
  d}^3p}{(2\pi)^3} \bar\Psi_V\left({\bf p}+\frac{2m_c}{E_V+M_V}{\bf
  \Delta} \right)\sqrt{\frac{\epsilon_q(\Delta)+
  m_q}{2\epsilon_q(\Delta)}}\cr
&&\times\Biggl\{\frac1{\epsilon_q(\Delta)}
\left[\frac{\epsilon_q(\Delta)-m_q}
{\epsilon_q(\Delta)+m_q}+\frac{M_V}{M_{B_c}}
\left(\frac{E_V+M_V}{\epsilon_q(\Delta)+m_q}+\frac{\epsilon_q(\Delta)-m_q}
{\epsilon_q(\Delta)+m_q}\right)\right]\cr
&&\times
\left[M_V-\epsilon_q\left(p+\frac{2m_c}{E_V+M_V}\Delta\right)
-\epsilon_c\left(p+\frac{2m_c}{E_V+M_V}\Delta \right)\right]\cr
&& +\frac{({\bf p\Delta})}{{\bf \Delta}^2}
\Bigglb(\Biggl(\frac1{2\epsilon_q(\Delta)}
\left[\frac{\epsilon_q(\Delta)-m_q}
{\epsilon_q(\Delta)+m_q}+\frac{M_V}{M_{B_c}}
\left(\frac{E_V+M_V}{\epsilon_q(\Delta)+m_q}+\frac{\epsilon_q(\Delta)-m_q}
{\epsilon_q(\Delta)+m_q}\right)\right]\cr
&&+\frac1{2m_b}\Biggl(\frac{\epsilon_q(\Delta)-m_q}{\epsilon_b(\Delta)+m_b}
+2\frac{M_V(E_V+M_V)}{[\epsilon_q(\Delta)+m_q][\epsilon_b(\Delta)+m_b]}
+\frac{M_V}{M_{B_c}}
\Biggl[\frac{\epsilon_q(\Delta)-m_q}{\epsilon_b(\Delta)+m_b}\cr
&&+
\frac{E_V+M_V}{\epsilon_b(\Delta)+m_b}
\left(1-\frac{2E_V}{\epsilon_q(\Delta)+m_q}\right)\Biggr]\Biggr)\Biggr)
\Biggl[M_{B_c}+M_V-\epsilon_b(p)-\epsilon_c(p)\cr
&&
-\epsilon_q\left(p+\frac{2m_c}{E_V+M_V}\Delta\right)
-\epsilon_c\left(p+\frac{2m_c}{E_V+M_V}\Delta \right)\Biggr]
+\frac{1}{2m_b}\Biggl(\frac{\epsilon_q(\Delta)-m_q}{\epsilon_q(\Delta)}\cr 
&& +
2\frac{M_V(E_V+M_V)}{\epsilon_q(\Delta)[\epsilon_q(\Delta)+m_q]}
+\frac{M_V}{M_{B_c}}
\Biggl[\frac{\epsilon_q(\Delta)-m_q}{\epsilon_q(\Delta)}\left(1-
\frac{E_V+M_V}{\epsilon_q(\Delta)+m_q}\right)\cr
&&-2\frac{E_V(E_V+M_V)}{\epsilon_q(\Delta)[\epsilon_q(\Delta)+m_q]}
\Biggr]\Biggr)
\Biggl[M_V-\epsilon_q\left(p+\frac{2m_c}{E_V+M_V}\Delta\right)\cr
&&
-\epsilon_c\left(p+\frac{2m_c}{E_V+M_V}\Delta \right)\Biggr]
\Biggrb)\Biggr\}\Psi_{B_c}({\bf p}), 
\end{eqnarray}

\begin{eqnarray}
  \label{eq:a2v}
  A_2^{V(2)}(q^2)&=&\frac{M_{B_c}+M_V}{2\sqrt{M_{B_c}M_V}}
\frac{2\sqrt{E_VM_V}}{E_V+M_V}\int \frac{{\rm
  d}^3p}{(2\pi)^3} \bar\Psi_V\left({\bf p}+\frac{2m_c}{E_V+M_V}{\bf
  \Delta} \right)\sqrt{\frac{\epsilon_q(\Delta)+
  m_q}{2\epsilon_q(\Delta)}}\cr 
&&\times
\frac{({\bf p\Delta})}{{\bf \Delta}^2}\frac{1}{2m_c}
\Biggl\{\frac{\epsilon_q(\Delta)-m_q}{\epsilon_q(\Delta)}\Bigglb(
-\frac{m_q}{\epsilon_q(\Delta)+m_q}\left(1+\frac{M_V}{M_{B_c}}\right)
[M_{B_c}-
\epsilon_b(p)-\epsilon_c(p)]\cr
&&+\left[1+\frac{M_V}{M_{B_c}}\left(1+
\frac{E_V+M_V}{\epsilon_q(\Delta)+m_q}\right)\right]
\Biggl[M_{B_c}-M_V-\epsilon_b(p)-\epsilon_c(p)\cr
&&
+\epsilon_q\left(p+\frac{2m_c}{E_V+M_V}\Delta\right)
+\epsilon_c\left(p+\frac{2m_c}{E_V+M_V}\Delta \right)\Biggr]\Biggrb)
\cr &&
+\Bigglb(2\frac{M_V(E_V+M_V)}
{[\epsilon_q(\Delta)+m_q][\epsilon_b(\Delta)+m_b]}\left(\frac{E_V}{M_V}
-1\right)-\left(
\frac{\epsilon_q(\Delta)+m_q}{\epsilon_b(\Delta)+m_b}
+\frac{m_q}{\epsilon_q(\Delta)}\right)\cr
&&\times\left[\frac{\epsilon_q(\Delta)-m_q}{\epsilon_q(\Delta)+m_q}
+\frac{M_V}{M_{B_c}}\left(
\frac{\epsilon_q(\Delta)-m_q}{\epsilon_q(\Delta)+m_q} 
+\frac{E_V+M_V}{\epsilon_q(\Delta)+m_q}\right)\right]
\Biggrb)
\Biggl[M_{B_c}+M_V-\epsilon_b(p)\cr
&&-\epsilon_c(p)
-\epsilon_q\left(p+\frac{2m_c}{E_V+M_V}\Delta\right)
-\epsilon_c\left(p+\frac{2m_c}{E_V+M_V}\Delta \right)\Biggr]
\Biggr\}\Psi_{B_c}({\bf p}), 
\end{eqnarray}

\begin{eqnarray}
  \label{eq:a01}
  A_0^{(1)}(q^2)&=&\sqrt{\frac{E_V}{M_V}}\int \frac{{\rm
  d}^3p}{(2\pi)^3} \bar\Psi_V\left({\bf p}+\frac{2m_c}{E_V+M_V}{\bf
  \Delta} \right)\sqrt{\frac{\epsilon_q(p+\Delta)+
  m_q}{2\epsilon_q(p+\Delta)}} \sqrt{\frac{\epsilon_b(p)+
  m_b}{2\epsilon_b(p)}}\cr
&&\times\Biggl\{1+\frac{M_{B_c}-E_V}{\epsilon_q(p+\Delta)+m_q}
\Bigglb(1+\Biggl[\frac{({\bf p\Delta})}{{\bf \Delta}^2}
-\frac23\frac{{\bf p}^2}{E_V+M_V}
\Biggl(\frac{1}{\epsilon_q(p+\Delta)+m_q}\cr
&&
+\frac{1}{\epsilon_c(p)+m_c}\Biggr)\Biggr]
\left[1+\frac1{2m_b}\left(\frac{{\bf \Delta}^2}{M_{B_c}-E_V}
+\epsilon_q(p+\Delta)+m_q\right)\right]\Biggrb)\cr
&&-\frac{{\bf p}^2}{6m_b[\epsilon_q(p+\Delta)+m_q]}\Biggr\}
\Psi_{B_c}({\bf p}),
\end{eqnarray}

\begin{eqnarray}
  \label{eq:a0s}
  A_0^{S(2)}(q^2)&=&\sqrt{\frac{E_V}{M_V}}\int \frac{{\rm
  d}^3p}{(2\pi)^3} \bar\Psi_V\left({\bf p}+\frac{2m_c}{E_V+M_V}{\bf
  \Delta} \right)\sqrt{\frac{\epsilon_q(\Delta)+
  m_q}{2\epsilon_q(\Delta)}}\Biggl\{-\frac{M_{B_c}-E_V}
{\epsilon_q(\Delta)+m_q}\frac1{\epsilon_q(\Delta)}\cr
&&\times\left(1-\frac{\epsilon_q(\Delta)-m_q}{M_{B_c}-E_V}\right)
\left[M_V-\epsilon_q\left(p+\frac{2m_c}{E_V+M_V}\Delta\right)
-\epsilon_c\left(p+\frac{2m_c}{E_V+M_V}\Delta \right)\right]\cr
&& +\frac{({\bf p\Delta})}{{\bf \Delta}^2}
\Bigglb((M_{B_c}-E_V)\Biggl[ 
\frac{1}{2\epsilon_q(\Delta)[\epsilon_q(\Delta)+m_q]}
\left(1+\frac{\epsilon_q(\Delta)-m_q}{M_{B_c}-E_V}\right)\cr
&&+\frac{1}{2m_b[\epsilon_b(\Delta)+m_b]}
\left(1-\frac{\epsilon_q(\Delta)-m_q}{M_{B_c}-E_V}\right)\Biggr]
\Biggl[M_{B_c}+M_V-\epsilon_b(p)-\epsilon_c(p)\cr
&&
-\epsilon_q\left(p+\frac{2m_c}{E_V+M_V}\Delta\right)
-\epsilon_c\left(p+\frac{2m_c}{E_V+M_V}\Delta \right)\Biggr]
-\frac{1}{2m_b}\frac{\epsilon_q(\Delta)-m_q}{\epsilon_q(\Delta)}\cr 
&&\times
\left(1+\frac{M_{B_c}-E_V}{\epsilon_q(\Delta)+m_q}\right)  
\Biggl[M_V-\epsilon_q\left(p+\frac{2m_c}{E_V+M_V}\Delta\right)
\cr&&
-\epsilon_c\left(p+\frac{2m_c}{E_V+M_V}\Delta \right)\Biggr]
\Biggrb)\Biggr\}\Psi_{B_c}({\bf p}), 
\end{eqnarray}

\begin{eqnarray}
  \label{eq:a0v}
  A_0^{V(2)}(q^2)&=&\sqrt{\frac{E_V}{M_V}}\int \frac{{\rm
  d}^3p}{(2\pi)^3} \bar\Psi_V\left({\bf p}+\frac{2m_c}{E_V+M_V}{\bf
  \Delta} \right)\sqrt{\frac{\epsilon_q(\Delta)+
  m_q}{2\epsilon_q(\Delta)}}
\frac{({\bf p\Delta})}{{\bf \Delta}^2}\frac1{2m_c}
\cr&&\times
\Biggl\{(M_{B_c}-E_V)
\Biggl[\frac{1}{\epsilon_b(\Delta)+m_b}
\left(1+\frac{\epsilon_q(\Delta)-m_q}{M_{B_c}-E_V}\right)
+\frac{m_q}{\epsilon_q(\Delta)[\epsilon_q(\Delta)+m_q]}\cr
&&\times
\left(1-\frac{\epsilon_q(\Delta)-m_q}{M_{B_c}-E_V}\right)\Biggr]
\Biggl[M_{B_c}+M_V-\epsilon_b(p)-\epsilon_c(p)
-\epsilon_q\left(p+\frac{2m_c}{E_V+M_V}\Delta\right)\cr
&&
-\epsilon_c\left(p+\frac{2m_c}{E_V+M_V}\Delta \right)\Biggr]
-\frac{\epsilon_q(\Delta)-m_q}{\epsilon_q(\Delta)}\Bigglb(M_{B_c}-
\epsilon_b(p)-\epsilon_c(p)\cr
&&-
\left(1-\frac{M_{B_c}-E_V}{\epsilon_q(\Delta)+m_q}\right)
\Biggl[M_{B_c}-M_V-
\epsilon_b(p)-\epsilon_c(p)
+\epsilon_q\left(p+\frac{2m_c}{E_V+M_V}\Delta\right)\cr
&&
+\epsilon_c\left(p+\frac{2m_c}{E_V+M_V}\Delta \right)\Biggr]
\Biggrb)\Biggr\}\Psi_{B_c}({\bf p}), 
\end{eqnarray}
where 
\[ \left|{\bf \Delta}\right|=\sqrt{\frac{(M_{B_c}^2+M_V^2-q^2)^2}
{4M_{B_c}^2}-M_V^2},\]
\[ E_V=\sqrt{M_V^2+{\bf \Delta}^2}. \]

\end{document}